\documentclass[%
    reprint,
    superscriptaddress,
    amsmath,amssymb,
    aps,
    noeprint, 
    prd,
]{revtex4-2}

\usepackage{graphicx}
\usepackage{dcolumn}
\usepackage{bm}
\usepackage{siunitx}
\usepackage[T5,T1]{fontenc}
\usepackage{hyperref}
\usepackage{booktabs} 

\newcommand{\dit}{d_{\mathrm{IT}}}
\newcommand{\etrunc}{E_{\mathrm{trunc}}}
\newcommand{\ineff}{\mathrm{inefficiency}}
\newcommand{\stoch}{\mathrm{stochasticity}}
\newcommand{\fleading}{f_{\mathrm{leading}}}
\newcommand{\qtot}{Q_{\mathrm{tot}}}

\newcommand{\eg}{\textit{e.g.}}
\newcommand{\ie}{\textit{i.e.}}

\newcommand{\pval}{p_{\mathrm{val}}}

\newcommand{\yb}{y_\mathrm{bundle}}
\newcommand{\yhatb}{\hat{y}_\mathrm{bundle}}
\newcommand{\Pb}{P_{\mathrm{bundle}}}
\newcommand{\Ps}{P_{\mathrm{single}}}
\newcommand{\Phatb}{\hat{P}_{\mathrm{bundle}}}
\newcommand{\Phats}{\hat{P}_{\mathrm{single}}}
\newcommand{\passmodel}{\mathcal{P_\mathrm{pass}}}

\newcommand{\KDEb}{\mathcal{T}_\mathrm{bundle}}
\newcommand{\KDEs}{\mathcal{T}_\mathrm{single}}
\newcommand{\KDEdata}{\mathcal{T}_\mathrm{data}}
\newcommand{\KDEtot}{\mathcal{T}_\mathrm{tot}}

\newcommand{\astroNorm}{\Phi_\mathrm{astro}}
\newcommand{\gammaAstro}{\gamma_\mathrm{astro}}
\newcommand{\cutoffEnergy}{E_{\mathrm{cutoff}}}
\newcommand{\domeff}{\epsilon_{\mathrm{DOM}}}
\newcommand{\iceabs}{\epsilon_{\mathrm{absorp}}}
\newcommand{\icescat}{\epsilon_{\mathrm{scat}}}
\newcommand{\convNorm}{\Phi_\mathrm{conv}}
\newcommand{\promptNorm}{\Phi_\mathrm{prompt}}
\newcommand{\deltaGamma}{\Delta\gamma_\mathrm{CR}}
\newcommand{\lambdaCR}{\lambda_\mathrm{CR}}

\newcommand{\muonNormNT}{\Phi^{\mathrm{NT}}_{\mathrm{atm}}}
\newcommand{\muonNormDPeV}{\Phi^{\mathrm{DPeV}}_{\mathrm{atm}}}

\begin{document}


\title{Probing the PeV Region in the Astrophysical Neutrino Spectrum using $\nu_\mu$ from the Southern Sky}


\affiliation{III. Physikalisches Institut, RWTH Aachen University, D-52056 Aachen, Germany}
\affiliation{Department of Physics, University of Adelaide, Adelaide, 5005, Australia}
\affiliation{Dept. of Physics and Astronomy, University of Alaska Anchorage, 3211 Providence Dr., Anchorage, AK 99508, USA}
\affiliation{Dept. of Physics, University of Texas at Arlington, 502 Yates St., Science Hall Rm 108, Box 19059, Arlington, TX 76019, USA}
\affiliation{School of Physics and Center for Relativistic Astrophysics, Georgia Institute of Technology, Atlanta, GA 30332, USA}
\affiliation{Dept. of Physics, Southern University, Baton Rouge, LA 70813, USA}
\affiliation{Dept. of Physics, University of California, Berkeley, CA 94720, USA}
\affiliation{Lawrence Berkeley National Laboratory, Berkeley, CA 94720, USA}
\affiliation{Institut f{\"u}r Physik, Humboldt-Universit{\"a}t zu Berlin, D-12489 Berlin, Germany}
\affiliation{Fakult{\"a}t f{\"u}r Physik {\&} Astronomie, Ruhr-Universit{\"a}t Bochum, D-44780 Bochum, Germany}
\affiliation{Universit{\'e} Libre de Bruxelles, Science Faculty CP230, B-1050 Brussels, Belgium}
\affiliation{Vrije Universiteit Brussel (VUB), Dienst ELEM, B-1050 Brussels, Belgium}
\affiliation{Dept. of Physics, Simon Fraser University, Burnaby, BC V5A 1S6, Canada}
\affiliation{Department of Physics and Laboratory for Particle Physics and Cosmology, Harvard University, Cambridge, MA 02138, USA}
\affiliation{Dept. of Physics, Massachusetts Institute of Technology, Cambridge, MA 02139, USA}
\affiliation{Dept. of Physics and The International Center for Hadron Astrophysics, Chiba University, Chiba 263-8522, Japan}
\affiliation{Department of Physics, Loyola University Chicago, Chicago, IL 60660, USA}
\affiliation{Dept. of Physics and Astronomy, University of Canterbury, Private Bag 4800, Christchurch, New Zealand}
\affiliation{Dept. of Physics, University of Maryland, College Park, MD 20742, USA}
\affiliation{Dept. of Astronomy, Ohio State University, Columbus, OH 43210, USA}
\affiliation{Dept. of Physics and Center for Cosmology and Astro-Particle Physics, Ohio State University, Columbus, OH 43210, USA}
\affiliation{Niels Bohr Institute, University of Copenhagen, DK-2100 Copenhagen, Denmark}
\affiliation{Dept. of Physics, TU Dortmund University, D-44221 Dortmund, Germany}
\affiliation{Dept. of Physics and Astronomy, Michigan State University, East Lansing, MI 48824, USA}
\affiliation{Dept. of Physics, University of Alberta, Edmonton, Alberta, T6G 2E1, Canada}
\affiliation{Erlangen Centre for Astroparticle Physics, Friedrich-Alexander-Universit{\"a}t Erlangen-N{\"u}rnberg, D-91058 Erlangen, Germany}
\affiliation{Physik-department, Technische Universit{\"a}t M{\"u}nchen, D-85748 Garching, Germany}
\affiliation{D{\'e}partement de physique nucl{\'e}aire et corpusculaire, Universit{\'e} de Gen{\`e}ve, CH-1211 Gen{\`e}ve, Switzerland}
\affiliation{Dept. of Physics and Astronomy, University of Gent, B-9000 Gent, Belgium}
\affiliation{Dept. of Physics and Astronomy, University of California, Irvine, CA 92697, USA}
\affiliation{Karlsruhe Institute of Technology, Institute for Astroparticle Physics, D-76021 Karlsruhe, Germany}
\affiliation{Karlsruhe Institute of Technology, Institute of Experimental Particle Physics, D-76021 Karlsruhe, Germany}
\affiliation{Dept. of Physics, Engineering Physics, and Astronomy, Queen's University, Kingston, ON K7L 3N6, Canada}
\affiliation{Department of Physics {\&} Astronomy, University of Nevada, Las Vegas, NV 89154, USA}
\affiliation{Nevada Center for Astrophysics, University of Nevada, Las Vegas, NV 89154, USA}
\affiliation{Dept. of Physics and Astronomy, University of Kansas, Lawrence, KS 66045, USA}
\affiliation{Centre for Cosmology, Particle Physics and Phenomenology - CP3, Universit{\'e} catholique de Louvain, Louvain-la-Neuve, Belgium}
\affiliation{Department of Physics, Mercer University, Macon, GA 31207-0001, USA}
\affiliation{Dept. of Astronomy, University of Wisconsin{\textemdash}Madison, Madison, WI 53706, USA}
\affiliation{Dept. of Physics and Wisconsin IceCube Particle Astrophysics Center, University of Wisconsin{\textemdash}Madison, Madison, WI 53706, USA}
\affiliation{Institute of Physics, University of Mainz, Staudinger Weg 7, D-55099 Mainz, Germany}
\affiliation{Department of Physics, Marquette University, Milwaukee, WI 53201, USA}
\affiliation{Institut f{\"u}r Kernphysik, Universit{\"a}t M{\"u}nster, D-48149 M{\"u}nster, Germany}
\affiliation{Bartol Research Institute and Dept. of Physics and Astronomy, University of Delaware, Newark, DE 19716, USA}
\affiliation{Dept. of Physics, Yale University, New Haven, CT 06520, USA}
\affiliation{Columbia Astrophysics and Nevis Laboratories, Columbia University, New York, NY 10027, USA}
\affiliation{Dept. of Physics, University of Oxford, Parks Road, Oxford OX1 3PU, United Kingdom}
\affiliation{Dipartimento di Fisica e Astronomia Galileo Galilei, Universit{\`a} Degli Studi di Padova, I-35122 Padova PD, Italy}
\affiliation{Dept. of Physics, Drexel University, 3141 Chestnut Street, Philadelphia, PA 19104, USA}
\affiliation{Physics Department, South Dakota School of Mines and Technology, Rapid City, SD 57701, USA}
\affiliation{Dept. of Physics, University of Wisconsin, River Falls, WI 54022, USA}
\affiliation{Dept. of Physics and Astronomy, University of Rochester, Rochester, NY 14627, USA}
\affiliation{Department of Physics and Astronomy, University of Utah, Salt Lake City, UT 84112, USA}
\affiliation{Dept. of Physics, Chung-Ang University, Seoul 06974, Republic of Korea}
\affiliation{Oskar Klein Centre and Dept. of Physics, Stockholm University, SE-10691 Stockholm, Sweden}
\affiliation{Dept. of Physics and Astronomy, Stony Brook University, Stony Brook, NY 11794-3800, USA}
\affiliation{Dept. of Physics, Sungkyunkwan University, Suwon 16419, Republic of Korea}
\affiliation{Institute of Basic Science, Sungkyunkwan University, Suwon 16419, Republic of Korea}
\affiliation{Institute of Physics, Academia Sinica, Taipei, 11529, Taiwan}
\affiliation{Dept. of Physics and Astronomy, University of Alabama, Tuscaloosa, AL 35487, USA}
\affiliation{Dept. of Astronomy and Astrophysics, Pennsylvania State University, University Park, PA 16802, USA}
\affiliation{Dept. of Physics, Pennsylvania State University, University Park, PA 16802, USA}
\affiliation{Dept. of Physics and Astronomy, Uppsala University, Box 516, SE-75120 Uppsala, Sweden}
\affiliation{Dept. of Physics, University of Wuppertal, D-42119 Wuppertal, Germany}
\affiliation{Deutsches Elektronen-Synchrotron DESY, Platanenallee 6, D-15738 Zeuthen, Germany}

\author{R. Abbasi}
\affiliation{Department of Physics, Loyola University Chicago, Chicago, IL 60660, USA}
\author{M. Ackermann}
\affiliation{Deutsches Elektronen-Synchrotron DESY, Platanenallee 6, D-15738 Zeuthen, Germany}
\author{J. Adams}
\affiliation{Dept. of Physics and Astronomy, University of Canterbury, Private Bag 4800, Christchurch, New Zealand}
\author{S. K. Agarwalla}
\thanks{also at Institute of Physics, Sachivalaya Marg, Sainik School Post, Bhubaneswar 751005, India}
\affiliation{Dept. of Physics and Wisconsin IceCube Particle Astrophysics Center, University of Wisconsin{\textemdash}Madison, Madison, WI 53706, USA}
\author{J. A. Aguilar}
\affiliation{Universit{\'e} Libre de Bruxelles, Science Faculty CP230, B-1050 Brussels, Belgium}
\author{M. Ahlers}
\affiliation{Niels Bohr Institute, University of Copenhagen, DK-2100 Copenhagen, Denmark}
\author{J.M. Alameddine}
\affiliation{Dept. of Physics, TU Dortmund University, D-44221 Dortmund, Germany}
\author{N. M. Amin}
\affiliation{Bartol Research Institute and Dept. of Physics and Astronomy, University of Delaware, Newark, DE 19716, USA}
\author{K. Andeen}
\affiliation{Department of Physics, Marquette University, Milwaukee, WI 53201, USA}
\author{C. Arg{\"u}elles}
\affiliation{Department of Physics and Laboratory for Particle Physics and Cosmology, Harvard University, Cambridge, MA 02138, USA}
% ...
\date{\today}

\collaboration{IceCube Collaboration}
\noaffiliation



\begin{abstract}

IceCube has observed a diffuse astrophysical neutrino flux over the energy region from a few TeV to a few PeV. At PeV energies, the spectral shape is not yet well measured due to the low statistics of the data. This analysis probes the gap between 1 PeV and 10 PeV by using high-energy downgoing muon neutrinos. To reject the large atmospheric muon background, two complementary techniques are combined. The first technique selects events with high stochasticity to reject atmospheric muon bundles whose stochastic energy losses are smoothed due to high muon multiplicity. The second technique vetoes atmospheric muons with the IceTop surface array. Using 9 years of data, we found two neutrino candidate events in the signal region, consistent with expectation from background, each with relatively high signal probabilities. A joint maximum likelihood estimation is performed using this sample and an independent 9.5-year sample of tracks to measure the neutrino spectrum. A likelihood ratio test is done to compare the single power-law (SPL) vs. SPL+cutoff hypothesis; the SPL+cutoff model is not significantly better than the SPL. High-energy astrophysical objects from four source catalogs are also checked around the direction of the two events. No significant coincidence was found. 

\end{abstract}

\maketitle

\section{Introduction}\label{sec:introduction}

The discovery of a diffuse flux of TeV to PeV high-energy astrophysical neutrinos by the IceCube Neutrino Observatory~\cite{Diffuse:2013:evidence-for-astro-neutrino, Diffuse:2013:two-pev-neutrino-events} in 2013 marked the beginning of high energy neutrino astronomy. Because neutrinos are expected to be produced at the same acceleration sites as cosmic rays through hadronic ($pp$ or $p\gamma$) interactions, measuring the spectrum of astrophysical neutrinos provides insights into the origin and acceleration mechanisms of cosmic rays. So far, only a few events have been observed with energies above \SI{1}{PeV}, hinting that the neutrino spectrum might exhibit a cutoff in the PeV region~\cite{Diffuse:2013:evidence-for-astro-neutrino, Diffuse:2014:3-year-HESE}. Such a cutoff is naturally expected if particles accelerated at astrophysical sites can only reach a finite maximum energy. This expectation is suggested by the spectral break observed in the ultra-high-energy cosmic ray spectrum, notably the ``second knee'' around \SI{100}{PeV}~\cite{Theory:2018:PDG}. For specific source classes, a cutoff in the neutrino spectrum could arise from various mechanisms, such as the diffusive escape of cosmic rays from sources~\cite{Theory:2013:test-hadronuclear-origin-of-pev-neutrinos, Flux:Senno-2015:cutoff-in-SFGs, Flux:Bartos-2015:SBGs, Flux:Fang-2018:AGN-CR-reservoir}, an energy cutoff in cosmic-ray sources~\cite{Flux:Tavecchio-2015:LLBLLac-pev-cutoff, Flux:Padovani-2015:BLLac}, or meson cooling in sources~\cite{Theory:2013:possible-origins-of-PeV-cutoff}. Measuring the cutoff energy can differentiate between these scenarios and provide insight into cosmic accelerators.

IceCube event topologies can generally be classified as cascades or tracks. Cascades are produced by charged-current (CC) interactions of $\nu_e$ and $\nu_\tau$, as well as by all-flavor neutral-current (NC) interactions, resulting in a nearly spherical light pattern. Muon tracks come from CC $\nu_\mu$ interactions and some $\nu_\tau$ interactions, if the $\tau$ decay includes a $\mu$. Decays of $W^\pm$ bosons produced at the Glashow resonance can also produce high-energy muons \cite{IceCube:2021rpz}. 

Previous IceCube analyses have targeted different neutrino energy ranges and flavors ($\nu_\mu$ vs. $\nu_e$ vs. $\nu_\tau$), but have not precisely characterized the spectral properties in the PeV region. The High Energy Starting Event (HESE) analyses~\cite{Diffuse:2013:evidence-for-astro-neutrino,Diffuse:2014:3-year-HESE, Diffuse:2021:7.5-year-HESE} looked for two signatures: muon tracks and cascades originating within the detector. These samples have high astrophysical purity, but the event selection has a relatively small effective volume. The Northern Tracks (NT) analyses~\cite{IceCube:2015qii,Diffuse:2016:6-year-northern-track, Diffuse:2022:9.5-year-diffuse-numu} look for energetic upgoing mostly-throughgoing tracks initiated by $\nu_\mu$ interactions outside the detector.  These muons can travel many kilometers, so the $\nu_\mu$ effective volume is much larger.  However, the sensitivity above \SI{1}{PeV} is limited because the Earth absorbs most high-energy neutrinos due to the increasing neutrino interaction cross sections~\cite{Model:2011:CSMS,IceCube:2017roe}. The Extremely High Energy (EHE) analyses~\cite{IceCube:2018fhm,Diffuse:2023:12-years-EHE} search for cosmogenic (GZK) neutrinos~\cite{Theory:1966:GZK-cutoff-proposal, Theory:1966:GZK-limit}, looking for neutrino-induced tracks or cascades, with a focus on energies above 10 PeV. 

To fill the gap between \SI{1}{PeV} and \SI{10}{PeV}, we developed a new event selection (referred to as DPeV in this paper) targeting downgoing, throughgoing high-energy tracks induced by astrophysical neutrinos~\cite{Yangdissertation}. It features a large effective volume and is not affected by Earth absorption. However, in this downgoing region, single and multiple muon tracks induced by cosmic rays dominate.

To reduce this background, two largely independent techniques are combined. The first technique selects tracks with more stochastic energy loss, \textit{i.e.}, with large fluctuations in specific energy loss, $dE/dx$. This preferentially selects single muons, rejecting more atmospheric muons which tend to come in multi-muon bundles where the energy loss fluctuations average out. The second technique uses the surface array of cosmic-ray detectors, IceTop~\cite{Det:2013:IceTop-paper}, to veto events that are accompanied by air showers. In the signal region, there is an irreducible background of single muons without accompanying IceTop hits, mostly created by energetic proton cosmic rays. This background is estimated with a data-driven method, and the selection cuts are chosen so that a 2:1 signal-to-background ratio is achieved in the final sample. 

The IceTop veto loses rejection power at sub-PeV primary energies. To provide an adequate lever arm (in energy) to fit the astrophysical neutrino energy spectrum, this sample is jointly fit with the 9.5-year Northern Track sample~\cite{Diffuse:2022:9.5-year-diffuse-numu} to measure the astrophysical neutrino spectrum. The likelihoods for a single-power-law (SPL) model and a SPL+cutoff model are compared. Because this DPeV sample has good angular resolution, checks for potential high-energy point sources near the direction of the events are performed.

The outline of the paper is as follows: Section~\ref{sec:event-selection} describes the event selection, including the stochasticity measurements and the IceTop veto. Section~\ref{sec:background-estimation} presents the background estimation method. Section~\ref{sec:systematic-uncertainty} summarizes the systematic uncertainties associated with the detector and the background estimation. Section~\ref{sec:analysis} discusses the analysis methods, while Sec.~\ref{sec:results} presents the results of the diffuse spectrum measurements and point source searches.

\section{Event Selection}\label{sec:event-selection}

\subsection{Detector}

The IceCube Neutrino Observatory is a cubic-kilometer neutrino detector located at the geographic South Pole~\cite{Det:2017:IceCube-paper,Halzen:2010yj}. The in-ice detector consists of 5,160 Digital Optical Modules (DOMs) deployed on 86 strings, with 60 DOMs on each string, positioned from \SI{1450}{m} to \SI{2450}{m} beneath the surface of the ice. Each DOM comprises a downward-facing photomultiplier tube (PMT) and digitization electronics. The average string spacing is around \SI{125}{m}, and the vertical spacing between adjacent DOMs is \SI{17}{m}. Eight of the 86 strings, with denser spacing of strings and DOMs, comprise the DeepCore subarray~\cite{Diffuse:2013:1-year-atm-nue-deepcore}. 

The air shower array, IceTop~\cite{Det:2013:IceTop-paper}, is located on the surface of the ice. It consists of 81 stations, each comprising two ice-filled tanks, located near one of the in-ice string locations. Each tank contains two DOMs operating at different gains. 

IceCube detects neutrinos through the Cherenkov photons produced by secondary charged particles created when neutrinos interact with the ice. A DOM records a ``hit'' when the PMT readout voltage corresponds to at least \SI{0.25}{photoelectrons} (PE)~\cite{Det:2009:DAQ-paper}. For the in-ice detector, a hit satisfies the hard-local-coincidence (HLC) criterion if its neighboring or next-to-neighboring DOMs are also hit within \SI{1}{\micro\second}; otherwise it is a soft-local-coincidence (SLC) hit~\cite{Det:2017:IceCube-paper}. IceTop detects particles from extensive air showers produced by cosmic ray interactions using ice-filled tanks. An HLC hit occurs when both tanks in a station are hit within \SI{1}{\micro\second}~\cite{Det:2013:IceTop-paper}. 

Monte Carlo simulated events were used to estimate both signals and backgrounds in the detector. Backgrounds come from atmospheric muons and from atmospheric neutrinos. For the simulations, the astrophysical neutrino signal was assumed to follow a per-flavor single power law (SPL) flux of $\Phi(E_\nu)= \Phi_{\text{astro}} \left(E_\nu/\SI{100}{TeV}\right)^{-\gamma}$, where $\Phi_{\text{astro}} = C_0\phi$ and $C_0= 10^{-18}\si{GeV^{-1}.cm^{-2}.s^{-1}.sr^{-1}}$, and $\gamma$ is the spectral index. Each neutrino flavor is assumed to have the same flux. We assume $\gamma=2$ in simulations; the specific choice does not matter because the muon background is estimated with a data-driven approach (see Section~\ref{sec:background-estimation}). The conventional (from pion and kaon decays) and prompt (from charmed meson decays) atmospheric neutrino energy-zenith angle distributions are based on the Matrix Cascade Equation solver (MCEq) software~\cite{Simulation:2015:MCEq}, assuming the Sibyll2.3c~\cite{Model:2019:sibyll2.3c} hadronic interaction model and the Gaisser-H4a~\cite{Model:2012:H3a-H4a} primary cosmic-ray composition model. Atmospheric muon events are simulated by CORSIKA~\cite{Simulation:1998:CORSIKA} also with 
Sibyll2.3c, but with the Gaisser-H3a flux~\cite{Model:2012:H3a-H4a}. 

All simulations use the PROPOSAL code \cite{Koehne:2013gpa} to propagate the muons through ice and then propagate the Cherenkov photons using the ice model SPICE-3.2.1~\cite{Thesis:2019:Martin-Rogen}.  SPICE~3.2.1 incorporates the glacial ice properties such as absorption, scattering, tilt of ice layers, and the anisotropy in light propagation~\cite{Det:2013:evidence-for-ice-anisotropy-SPICELea}.

\subsection{Preselections and quality cuts}

The data used for this analysis were taken with the full detector configuration (IC86) over nine years from May 2011 to May 2020. The ``Pass2'' re-calibration campaign~\cite{Det:2020:pass-2-campaign}, which features a more accurate charge response compared to earlier calibrations, has been applied to all data. To ensure data quality, only data where all eighty-six in-ice strings and at least 310 IceTop DOMs were active are used. This analysis follows a blind procedure, where 10\% of the full data, referred to as the ``burn sample'', are only used during the cut optimization 
and to design the muon background estimation procedure. No burn sample events are found in the signal region.
The burn sample is discarded, and the live time of the unblinded data used for obtaining the physics results is 2,792.5 days.

The event selection is summarized in Table \ref{tab:select}. It first applies a series of cuts for data cleaning and preselections. DeepCore strings are removed to ensure a uniform detector geometry. 
Only in-ice HLC hits are kept, and a sliding time window of width \SI{6000}{ns} is applied to the distribution of collected photoelectrons over time to identify the signal peak. Random hits outside the identified time range are then removed.
The event directions are reconstructed using the SplineMPE algorithm~\cite{Det:2014:splineMPE-thesis}. A cut on the reconstructed zenith angle, $\theta_{\mathrm{reco}} < 90^\circ$, is applied to select downgoing events. The muon energy, $\etrunc$, is reconstructed using Truncated Energy algorithm (bins method)~\cite{Reco:2013:truncated-energy}, which calculates $dE/dx$ values using 120-meter bins along the muon track. To reduce bias from large stochastic losses, the top 40\% of bins with the largest energy losses are removed, and the average $\langle dE/dx \rangle_{\mathrm{trunc}}$ is used to compute the muon energy near the center of the detector. An energy cut of $\etrunc > \SI{300}{TeV}$ and a total charge cut (summed photoelectrons over all in-ice DOMs) of $\qtot > \SI{4000}{PE}$ are applied to select high-energy events.

\begin{table}
    \centering
    \begin{tabular}{|l|l|}
    \hline
    \multicolumn{2}{c}{Preselection/quality cuts} \\
    \hline
         Spline MPE & $\theta_{\rm reco} <90^0$ \\
         Truncated Energy & $E_{\rm reco} > 300$ TeV\\
         Event charge & $Q_{\rm tot} > 4000$ pe\\
         Track length & $>650$ m \\
         Truncated Energy & $N_{\rm bins} >6$  \\
         Track DCA & $<500$ m from ctr.of IceCube \\
         Number of clusters & $=1$ \\
         Not a cascade & Cascade Likelihood $<19$ \\
         \hline
    \end{tabular}
    \caption{Summary of the preselection and data quality cuts.}
    \label{tab:select}
\end{table}

After the preselections, our sample contains well-reconstructed, high-energy, downgoing tracks with reliable IceTop information. We further apply a series of quality cuts. The track length within the instrumented volume, defined as the projected distance between the first-hit DOM and the last-hit DOM, is required to be larger than \SI{650}{m}. Additionally, the number of \SI{120}{m}-long bins from Truncated Energy~\cite{Reco:2013:truncated-energy} before truncation is required to be $\geq 6$. Furthermore, the distance from the track to the center of the IceCube array is required to be $< \SI{500}{m}$. These three cuts select long and well-contained muon tracks such that stochasticity measurements and energy reconstruction are reliable. To remove coincident events (events containing two or more independent muons), the number of clusters of causally connected hits (hits separated by a distance and time consistent with the speed of light in ice) in an event~\cite{Reco:2012:TopologicalSplitter} is required to be 1. Events were fit to both cascade and track hypotheses. The likelihood of the event under the cascade hypothesis~\cite{Diffuse:2011:5-yr-cascade-AMANDA} is required to be $<19$. These two cuts reduce the fraction of coincident events to approximately 2\% of the total event sample.

\subsection{Stochasticity measurement}

To reduce the background, we calculate the stochasticity of a muon track and reject events with small variations in energy deposition. This method works because high-energy muons exhibit stochastic energy losses in the ice due to bremsstrahlung, pair production, and photonuclear interactions~\cite{Theory:2018:PDG}. For single muons, the energy loss pattern along the track shows large fluctuations. In contrast, multi-muon bundles, commonly produced by air showers, exhibit smoother energy-loss patterns, as individual fluctuations average out. Specifically, individual muons in the bundle have lower energies due to the large multiplicity, leading to less stochastic energy losses. Furthermore, in the regime where stochastic losses dominate (when each muon has an energy above \SI{1}{TeV}), the average energy loss for a muon bundle is the same as that of a single muon with the same energy. However, fluctuations in the overall energy loss profile decrease as the number of muons in the bundle increases. The stochasticity variable is designed to quantify the fluctuations in the muon energy loss pattern along the track.

To measure stochasticity for an event, the specific energy loss ($dE/dx$) along the track is measured in \SI{120}{m} long bins~\cite{Reco:2013:truncated-energy}, with each bin $i$ having measured energy loss $y_i$. A linear function $\hat{y}$ is fitted to the measured muon energy losses. The reduced $\chi^2$ value of the fit is then used as the stochasticity value

\begin{equation}
    \text{stochasticity} = \log_{10}\left( \frac{1}{\mathrm{d.o.f}}\sum_{i}^{\text{bins}}\frac{(y_i-\hat{y_i})^2}{y_i}\right)
\end{equation}
where $\mathrm{d.o.f}$ is the number of degrees of freedom.
The denominator is chosen to increase sensitivity to outliers in the distribution of $y_i$. This formula is compared to several methods from previous analyses~\cite{CR:2020:IceTop-veto-results, CR:2016:characterize-atm-muon-flux-in-IC}, and it best separates the distributions of simulated signal and background events.

\begin{figure}
    \centering
    \includegraphics[width=0.99\linewidth]{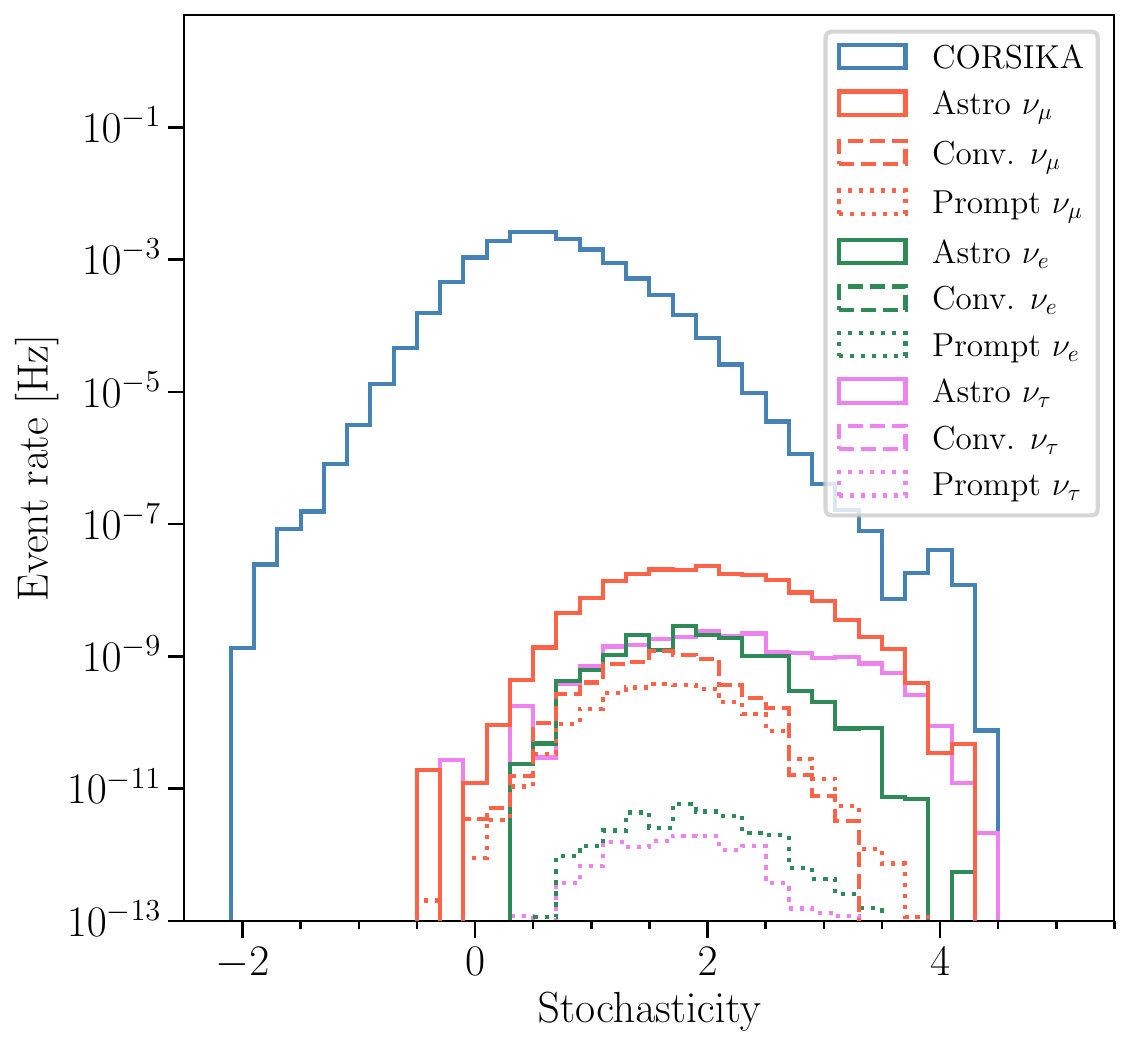}
    \caption{The stochasticity distribution for astrophysical (solid), conventional (dashed) and prompt (dotted) atmospheric neutrinos and atmospheric muons (CORSIKA, blue) after the preselections and the quality cuts. An $E^{-2}$ spectrum is assumed for astrophysical neutrinos. Most of the $\nu_e$ events that survive these cuts are from the Glashow resonance, with the $W^\pm$ decaying to a muon plus neutrino.}
    \label{fig:stochasticity}
\end{figure}

The stochasticity distributions of all the simulated components after preselections and quality cuts are shown in Fig.~\ref{fig:stochasticity}. Astrophysical neutrinos have a relatively hard spectrum, which results in distributions distinct from those of atmospheric muons. Because more energetic muons exhibit more stochastic energy losses, the measured stochasticity should be slightly positively correlated with the reconstructed muon energies. So, atmospheric neutrinos generally have smaller stochasticities. The signal region in this analysis uses a cut $\stoch > 3.0$ to remove most of the background. The rationale for choosing this cut value will be discussed in Section~\ref{sec:background-estimation}.

\subsection{IceTop veto and inefficiency}\label{subsec:icetop-veto-and-inefficiency}

\begin{figure}
    \centering
    \includegraphics[width=0.99\linewidth]{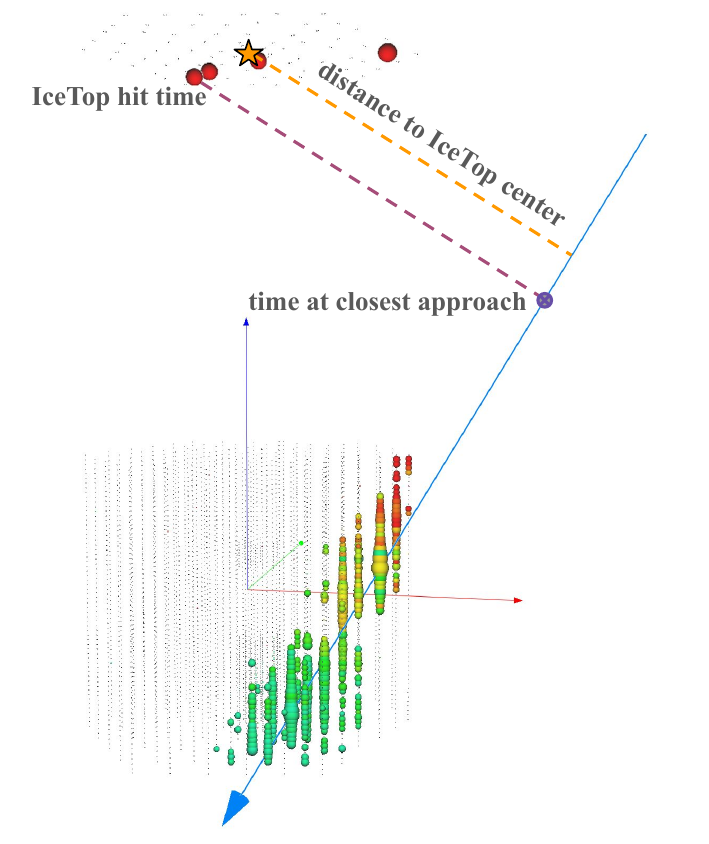}
    \caption{The geometry of the IceTop veto mechanism. The IceTop inefficiency is first parameterized using the muon energy and the muon track's distance to the center of IceTop (orange line).}
    \label{fig:idea_of_veto}
\end{figure}

The IceTop veto removes events accompanied by an air shower, which may create HLC or SLC (mostly from GeV single muons) IceTop tank hits~\cite{Det:2013:IceTop-paper}.  The concept is shown in Fig.~\ref{fig:idea_of_veto}.  For each IceTop hit at time $t_i$, the time for the muon at its closest approach position (purple dot) to the hit IceTop tank is calculated as $t_c$. If $\Delta t = t_i - t_c$ falls within a time window of $[-\SI{700}{ns}, \SI{1700}{ns}]$, the hit is considered correlated with the muon track. Events with more than one correlated IceTop hit are vetoed. The two-hit veto threshold is chosen to maximize veto power while maintaining a relatively low false positive rate. The time window is selected to capture the peak of the $\Delta t$ distribution of data, which has a long tail at positive values due to the shower curvature~\cite{Det:2013:IceTop-paper}.

To determine the probability of vetoing real astrophysical neutrino events, an equal-length background window is chosen at $[-\SI{4500}{ns}, \SI{-2100}{ns}]$. The probability of vetoing astrophysical neutrinos due to random hits is found to be 10\%, with negligible uncertainty, independent of energy and zenith angle, so astrophysical neutrino MC events are reweighted by a factor of 0.9.

\begin{figure}
    \centering
    \includegraphics[width=0.99\linewidth]{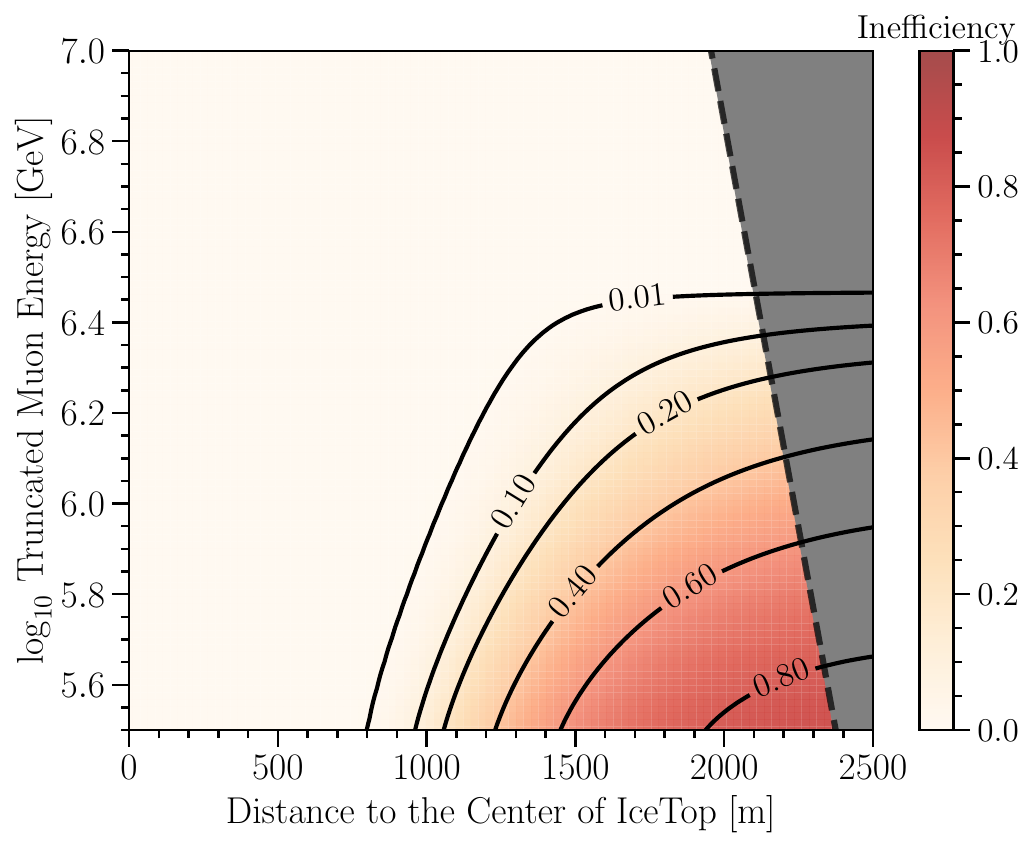}
    \caption{The baseline (9-year average) IceTop inefficiency model, parametrized using the full data in the low stochasticity region. The gray band to the right of the dashed line is the low-statistics region and is removed from this analysis. The contours are for the given estimated IceTop inefficiency, as a function of distance to the center of IceTop and estimated muon energy. The final cut is the line with an inefficiency of 0.01.}
    \label{fig:inefficiency_model}
\end{figure}

The IceTop veto is very efficient for air showers that pass near the center of IceTop. The determination of the IceTop inefficiency model (where a smaller inefficiency indicates a more effective veto) is discussed in Appendix~\ref{app:icetop}. First, the IceTop inefficiency of the data is calculated using low-stochasticity events that contain very little signal. A parametrization of the inefficiency is then fitted to the data, with constraints on the model parameters applied to mitigate the impact of statistical uncertainties. The model also accounts for the gradual decline in efficiency over time due to snow buildup on IceTop.

The baseline inefficiency model is shown in Fig.~\ref{fig:inefficiency_model}, as a function of muon energy and distance to the center of IceTop. The contours are lines corresponding to a given minimum inefficiency. If a maximum inefficiency is chosen, then IceTop can provide useful information for events to the right of and above the corresponding line. Efficiency decreases with increasing distance from IceTop and decreasing shower energy. The region to the right of the black dashed line has almost no data and is excluded from the analysis. Although statistics are limited for $\log_{10}\etrunc/{\rm GeV} > 6.5$, higher-energy muons should have larger average shower energies, resulting in a lower inefficiency at higher energies; therefore, the model retains its validity at high energy.

The signal region includes a cut of $\ineff < 0.01$ applied to data, selecting events in the veto-effective region. This cut is also applied to MC events but using the baseline  (9-year average) inefficiency. This approach does not  introduce data-MC discrepancies. 
Since the yearly modeling does not capture inefficiency variations within a year, this potential uncertainty is later treated as a systematic effect.

Several IceCube studies have used the IceTop array as a veto for downgoing tracks. The EHE analyses remove tracks with two or more correlated IceTop hits~\cite{Diffuse:2016:6-year-EHE, Diffuse:2023:12-years-EHE}, but they do not require detailed background estimation due to the low atmospheric muon expectations above \SI{10}{PeV}. Realtime neutrino alert streams, triggered by high-energy tracks, use IceTop to tag obvious background events~\cite{PS:2017:realtime-alert-system, PS:2023:IceCat-1}. More sophisticated veto methods that use both the time and the charges of IceTop hits are shown to be effective in the near-vertical region, but a background estimation is not performed due to the lack of combined in-ice and IceTop simulations~\cite{CR:2017:IceTop-veto-perforamnce, CR:2020:IceTop-veto-results}. Here, we demonstrate the application of the IceTop veto to the entire downgoing region, combined with the stochasticity variable and a robust data-driven background estimation method.

\subsection{Energy-zenith phase space cut}

\begin{figure}
    \centering
    \includegraphics[width=0.99\linewidth]{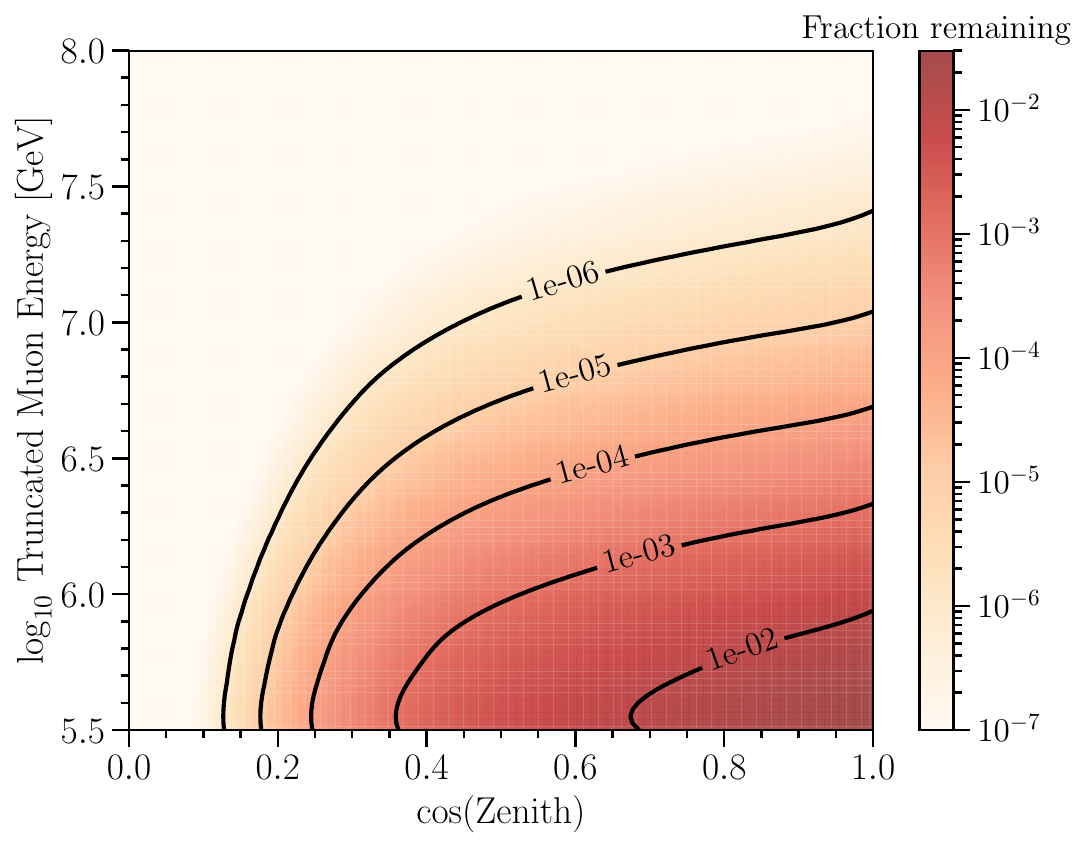}
    \caption{Contours used for the energy-zenith phase space cut. Each contour value represents the fraction of CORSIKA events that remains after the exclusion of all events below the contour. The final cut selects events with a phase space contour value less than 0.01.}
    \label{fig:phase_cut}
\end{figure}

After the stochasticity cut and IceTop veto, the atmospheric muon background is reduced by applying a cut in the energy-zenith phase space. Astrophysical neutrinos are isotropic in zenith with a harder spectrum, while atmospheric muons cluster in the vertical direction with a softer spectrum. Therefore, removing events in the low-energy, near-vertical region improves the signal-to-background ratio. This analysis excludes events below a contour obtained from the CORSIKA atmospheric muon distribution, shown in Fig.~\ref{fig:phase_cut}. Contour values indicate the fraction of the remaining CORSIKA events if events below the contour line are removed. In this paper, this cut is described as phase space < $\beta$, where events below the contour with a value of $\beta$ are removed. The optimal cut follows the contour for phase space $ < 0.01$ in the figure, reducing the atmospheric muon background by a factor of $\sim O(10^2)$.

\section{Background estimation and signal region}\label{sec:background-estimation}

The main background is from single muons and bundles of muons from cosmic-ray air showers. A data-driven model for this background is presented in Appendix \ref{app:muons}. Briefly, it divides the background into single muons and multiple muons, and extracts stochasticity distributions for both, using IceTop as a tag to remove any signal from the sample. The model is used to determine the fraction of single and multiple muons, and then to calculate the background -- estimated as $1.97^{+0.12}_{-0.14}$ events (the uncertainty quoted here is systematic only). The same method also estimates the background distribution in energy and zenith angle. This distribution is necessary for estimating the signalness, defined as $S/(S+B)$, of the observed data events. 

With the IceTop veto, the secondary background, from atmospheric neutrinos, is negligible. The atmospheric-neutrino flux is at least an order of magnitude lower than astrophysical neutrino fluxes after the preselections and quality cuts, mainly due to the high-energy cuts applied. Additionally, atmospheric neutrinos are accompanied by muon bundles from the same air shower, leading to a smaller observed stochasticity distribution than that shown in Fig.~\ref{fig:stochasticity}, which is based on neutrino simulations without accompanying bundles. Finally, atmospheric neutrinos are accompanied by air showers, allowing them to be vetoed by IceTop. The remaining atmospheric neutrino background is small enough that it can be neglected.

The signal region was determined by roughly optimizing the cuts on stochasticity, IceTop inefficiency and phase space, subject to the constraints that the expected signal:background ratio ($S:B$) for each event was at least 2.0, and that enough data events remained after the IceTop inefficiency cut to allow for the stochasticity curves to be determined. The optimization was performed using the burn sample, where $S$ is estimated using simulated neutrino events, and $B$ is obtained from the data-driven background modeling. The minimum $S:B$ ratio was chosen to reduce the effects of systematic uncertainties and provide a reasonably-pure sample for point-source searches.

The final cuts were $\stoch > 3.0$, $\ineff < 0.01$, the phase space contour $< 0.01$, and passing the IceTop veto. After unblinding, two events passed these cuts, in good agreement with the background estimate of $1.97^{+0.12}_{-0.14}$ events.

\section{Systematic Uncertainties}\label{sec:systematic-uncertainty}

\subsection{In-Ice Detector systematic uncertainties}

The in-ice detector systematics are associated with imperfect knowledge about the detector's responses and ice properties. The main uncertainties are from the optical efficiency of DOMs~\cite{Thesis:2019:measure-domeff-using-minimum-ionizing-muons}, the absorption and scattering of light in the bulk ice~\cite{Det:2006:optical-property-of-south-pole-ice}, and the scattering of the ``hole ice''~\cite{Det:2016:hole-ice-measurement}, which is formed after the drill holes refreeze~\cite{Det:2017:IceCube-paper}. These uncertainties are parameterized as nuisance parameters that affect the neutrino distributions in the signal region. The detector systematics for the atmospheric muon background are not parameterized, due to the lack of dedicated simulation datasets. The detector systematics have relatively small effects on the background normalization and shape due to the data-driven background estimation approach. Most importantly, statistical uncertainties in data dominate during subsequent physics measurements due to the very small sample size of downgoing events.

The uncertainty in the optical efficiency of DOMs is parameterized by $\domeff$, scaling the efficiency of all in-ice DOMs. Calibration studies with minimum-ionizing muons indicate a 10\% uncertainty in this parameter~\cite{Reco:2014:energy-reco-paper}.  Dedicated neutrino simulation datasets are generated with $\domeff$ settings varying within $\pm10\%$ around the nominal value of 1, and the effects due to $\domeff$ on distributions in the observable (muon-energy and zenith-angle) space are parametrized using a spline interpolation~\cite{IceCube:2023asu,IceCube:2024csv}. This method applies to all detector systematics parameterizations. A higher $\domeff$ generally leads to more events at higher energies. 

The bulk (glacial) ice contains impurities that affect the propagation of photons~\cite{Det:2006:optical-property-of-south-pole-ice}, whose effect has been incorporated into the ice model (\ie, SPICE-3.2.1 for this analysis)~\cite{Thesis:2019:Martin-Rogen}. Measurements of depth-dependent absorption and scattering coefficients, $\iceabs$ and $\icescat$, have uncertainties from 5\% to 10\%, derived from the LED flasher studies~\cite{Det:2023:ice-layer-undulation-mapping-FTPv3, Det:2013:measure-ice-property-SPICEMie}. Uncertainties due to $\icescat$ have negligible effect on the event distributions, while the systematics of $\iceabs$ are parameterized in the fit. An increase of $\iceabs$ leads to fewer events at higher energies.

The above systematics pertain to the bulk ice. During the refreezing of drill holes, air bubbles are pushed towards the center~\cite{Det:2016:hole-ice-measurement}. The result is that the hole ice has significantly more scattering. This is simulated via a modified DOM angular acceptance. More hole-ice scattering decreases the PMT acceptance in the forward direction and increases the acceptance in the backward direction. The angular acceptance is parameterized by a two-parameter ($p_0$ and $p_1$) function defined in Ref.~\cite{Reco:2023:unified-hole-ice-p0-p1}; $p_0$ primarily affects the forward acceptance (higher $p_0$ increases it), while $p_1$'s effect is orthogonal to $p_0$'s. The effect of $p_1$ on the downgoing sample is found to be degenerate with that of $p_0$ and is therefore ignored.

\subsection{Muon background estimation uncertainties}\label{subsec:muon-bg-estimation-uncertainties}

The muon background uncertainties are due to uncertainties on the background model, including in the stochasticity templates for single and multiple muons, and the fraction of events that are single muons.  Uncertainty on the latter comes mainly from the uncertainty in the fraction of cosmic-rays that are protons, since protons are most likely to produce single muon air showers.  These uncertainties are estimated by using Monte Carlo simulations with different compositions.  The muon background uncertainties are discussed in detail in Appendix \ref{app:muonuncertainties}. 

\subsection{Systematic uncertainties of the Northern Tracks sample}

The downgoing sample is jointly fit with the 9.5-year Northern Tracks sample~\cite{Diffuse:2022:9.5-year-diffuse-numu}. The Northern Tracks sample targets upgoing throughgoing tracks traveling through the Earth and has a sensitive range from \SI{15}{TeV} to \SI{5}{PeV}. A zenith cut of $\theta > 85^\circ$ selects upwardgoing and near-horizontal events in a region with little muon background. There is no overlap between data and simuation from the two samples. 

Since Ref.~\cite{Diffuse:2022:9.5-year-diffuse-numu}, some updates have been made to the data and the corresponding simulations. The IC59 data~\cite{Diffuse:2014:northern-tracks-IC59}, where the detector ran in the 59-string configuration, were dropped, leaving 651,377 events remaining with a live time of 8.18 years~\cite{Diffuse:2022:9.5-year-diffuse-numu}. The astrophysical $\nu_\tau$ flux is now included in the simulation, while the original analysis used reweighted $\nu_\mu$ simulation. An updated MCEq~\cite{Simulation:2015:MCEq} is used to calculate atmospheric neutrino fluxes and a special MC simulation dataset, SnowStorm, is used to better incorporate detector systematics~\cite{Stats:2019:SnowStorm-method}. These updates do not lead to significant shifts in the neutrino spectrum. 

Most of the systematic uncertainties are incorporated as nuisance parameters in the fit.   The Northern Tracks sample is dominated by atmospheric neutrinos, which are parameterized following Ref.~\cite{Diffuse:2022:9.5-year-diffuse-numu}. $\Phi_{\mathrm{conv}}$ and $\Phi_{\mathrm{prompt}}$ are the normalizations of the conventional and prompt atmospheric neutrino fluxes, respectively. The shape uncertainties of the conventional flux are characterized by a set of parameters (${\rm Atm}_H$, ${\rm Atm}_W$, ${\rm Atm}_Y$, ${\rm Atm}_Z$), which account for uncertainties in the pion and kaon productions~\cite{Model:2006:Barr-parameters}. $\Delta \gamma_{\mathrm{CR}}$ accounts for the shape uncertainty of the primary cosmic ray flux, which reweights the atmospheric neutrino fluxes to a different spectral index, leaving the overall normalization roughly unchanged. The assumptions of the primary flux models also affect the atmospheric neutrino flux, which is parameterized by $\lambda_{\mathrm{CR}}$ that interpolates between the H3a~\cite{Model:2012:H3a-H4a} and GST-4~\cite{Model:2013:GST} models. $\Phi^{\mathrm{NT}}_{\mathrm{atm}}$ is the normalization of the atmospheric muon template, which mostly consists of near-horizontal atmospheric muons.

\subsection{Summary of nuisance parameters}\label{subsec:summary-of-nuisance-parameters}

\begin{table}[htbp]
    \centering
    \label{tab:nuisance_parameters}
        \begin{tabular}{l||c|c|c|c|cc}
            \hline
            Syst. & Nominal & Prior & Width & Bounds & Sample \\ 
            \hline
            $\domeff$ & 1.0 & Flat & - & [0.9, 1.1] & both \\
            $\iceabs$ & 1.0 & Flat & - & [0.9, 1.1] & both \\
            $\icescat$ & 1.0 & Flat & - & [0.9, 1.1] & NT \\
            $p_0$ & -0.27 & Flat & - & [-0.84, 0.3] & both \\
            \hline
            $\convNorm$ & 1.0 & Flat & - & [0, $\infty$) & NT \\
            $\promptNorm$ & 1.0 & Flat & - & [0, 2.5] & NT \\
            $\deltaGamma$ & 0.0 & Flat & - & [-1, 1] & NT \\
            $\lambdaCR$ & 0.0 & Gauss & 1.0 & [-1, 2] & NT \\
            ${\rm Atm}_H$ & 0.0 & Gauss & 0.15 & [-0.5, 0.5] & NT \\
            ${\rm Atm}_W$ & 0.0 & Gauss & 0.40 & [-0.5, 0.5] & NT \\
            ${\rm Atm}_Y$ & 0.0 & Gauss & 0.30 & [-0.5, 0.5] & NT \\
            ${\rm Atm}_Z$ & 0.0 & Gauss & 0.12 & [-0.5, 0.5] & NT \\
            $\muonNormNT$ & 1.0 & Gauss & 0.05 & [0, $\infty$) & NT \\
            $\muonNormDPeV$ & 1.97 & Gauss & 1.40 & [0, 4.79] & DPeV \\ 
            \hline
        \end{tabular}
    \caption{Nuisance parameters used in the joint fit. Each parameter is either used in the Northern Tracks (NT) sample, or the downgoing tracks (DPeV) sample, or both. Parameters shared by both samples indicate a full correlation by construction: any changes in a shared parameter simultaneously affect the MC expectation in both samples.}\label{table:nuisance-params}
\end{table}

The nuisance parameters used in the joint fit are listed in Table~\ref{table:nuisance-params}. The motivations for the choices of priors and bounds for atmospheric-neutrino related parameters are described in Ref.~\cite{Diffuse:2022:9.5-year-diffuse-numu}. If a nuisance parameter is used for both samples, it is assigned the same value for each sample, simultaneously affecting the corresponding MC distributions of the two samples.

The bounds on detector systematics are obtained from calibration studies as mentioned before, which typically assume a 10\% uncertainty. A subtlety arises in the parameterization of the hole-ice $p_0$ parameter for the downgoing tracks sample, which is based on simulations that assume a baseline value of 0. In contrast, the SnowStorm simulation assumes a nominal value of $-0.27$, derived from a calibration study using the LED flasher data~\cite{Reco:2023:unified-hole-ice-p0-p1,Thesis:2021:Joeran-diffuse-numu}. Although the sets are generated around different central values, the best fit $p_0$ is in the valid range for both systematics parametrizations.

The prompt normalization is assumed to have an upper bound of 2.5 times the MCEq prediction ~\cite{Simulation:2015:MCEq}, which is numerically similar to the 2008 ERS calculation ~\cite{Model:2008:ERS}.  This assumption is motivated by the 90\% confidence level upper limit on the prompt flux normalization, of $5 \times \Phi_{\mathrm{BERSS}}$ found in the IceCube 6-year cascade analysis~\cite{Diffuse:2020:6-year-cascades}. The BERSS prompt flux model~\cite{Model:2015:BERSS} updates the ERS prompt flux model~\cite{Model:2008:ERS}, incorporating data from RHIC and LHC. The BERSS normalization is about half that of the ERS calculation at all energies. 

\section{Analysis method}
\label{sec:analysis}

\subsection{Binned effective likelihood}

In this analysis, measurements of the astrophysical neutrino spectrum are performed using a joint binned maximum likelihood estimation~\cite{Diffuse:2015:first-combined-fit}. In each bin $i$ of the observables, the likelihood $\mathcal{L}(\vec{\Theta}|D_i)$ depends on the physics parameters $\vec{\theta}$ and the nuisance parameters $\vec{\eta}$, collectively denoted as $\vec{\Theta}$, and the observed data $D_i$. To account for the limited MC statistics, particularly for the downgoing sample, an effective Schneider-Argüelles-Yuan (SAY) likelihood~\cite{Stats:2019:SAYLLH} is used (method $L_{\rm eff}$). It has improved coverage and is computationally efficient. When MC statistics are limited, a widening of the confidence interval is anticipated compared to the interval obtained using the Poisson likelihood. This improved coverage is especially important for the downgoing track sample because of the relatively small number of MC events per bin in the observable space (see Section~\ref{sec:non-uniform-binning}).

The data and MC simulations are binned in the two-dimensional observable space of $\etrunc$ and $\cos\theta$, where the signal and background distributions differ. The downgoing and Northern Tracks samples are binned separately. The joint likelihood is:
\begin{equation}
    \begin{aligned}
        \mathcal{L}^{\text{joint}}_{\text{SAY}}(\vec{\Theta}|D) &= 
        \prod\limits_{\text{bin m}}^M \mathcal{L}^{\text{NT}}_{\text{SAY}}(\vec{\Theta}|D_m) 
        \times
        \prod\limits_{\text{bin n}}^N \mathcal{L}^{\text{DPeV}}_{\text{SAY}}(\vec{\Theta}|D_n) \\
        &\times \quad \prod\limits_{\text{nuisance j}} \exp\left( -\frac{1}{2} \left(\frac{\eta_{j} - \eta_{j,0}}{\sigma[\eta_{j}]} \right)^2\right).
    \end{aligned}
\end{equation}
The first term compares the product of likelihoods for the Northern Tracks sample ($\mathcal L_{\text SAY}^{\text NT}$) and downgoing sample ($\mathcal L_{\text SAY}^{\text DPeV}$), while the second term accounts for the likelihoods of changes in the nuisance parameters.  
Each nuisance parameter $\eta_j$ has a nominal value is $\eta_{j,0}$ and  prior width $\sigma[\eta_j]$.  For nuisance parameters with flat priors, the second (exponential) term is not included.  The L-BFGS-B algorithm~\cite{Stats:1989:L-BFGS-B} is used to maximize the log-likelihood to find the best-fit parameters. Confidence intervals are obtained using the profile likelihood method employing Wilks' theorem~\cite{Stats:1938:wilks-theorem}.

\subsection{Non-uniform binning}
\label{sec:non-uniform-binning}

The Northern Tracks sample is binned in 50 logarithmically spaced bins in reconstructed muon energy, from \SI{100}{GeV} to \SI{10}{PeV}, and 33 bins in $\cos\theta$ from -1 to 0.0872, chosen to cover the full range of zenith angles where neutrinos dominate. The DPeV sample covers the region $\cos\theta \in [0.0872, 1]$ and $\etrunc \in [\SI{e6}{GeV}, \SI{e9}{GeV}]$. 
Due to limited MC statistics for the downgoing sample, uniform binning results in many empty bins, leading to undefined log-likelihood values. To avoid this, a non-uniform binning is used, determined by the decision boundaries of a trained regression tree~\cite{Diffuse:2023:non-uniform-binning}. The training data consists of a combination of simulated neutrino and CORSIKA events in the signal region. The tree uses the two observables to predict $y$, a linear combination of zenith and reconstructed energy:
\begin{align}
    \widetilde{\Theta} &= \frac{\cos\theta_{\mathrm{true}} - \langle \cos\theta_{\mathrm{true}} \rangle}{\sigma(\cos\theta_{\mathrm{true}})},\\
    \widetilde{\mathrm{E}} &= \frac{\log_{10}E_{\mathrm{trunc}} - \langle \log_{10}E_{\mathrm{trunc}} \rangle}{\sigma(\log_{10}E_{\mathrm{trunc}})} \\
    y &= \alpha \cdot \widetilde{\Theta} + \beta \cdot \widetilde{\mathrm{E}}.
\end{align}
During training, each split minimizes the mean squared error within the resulting child nodes, with the final decision boundaries defining the bins. The minimum number of MC events in each bin (\textit{i.e.}, at each leaf node) is set to two, ensuring that the bins are non-empty. The resultant binning is relatively uniform because the target variable $y$ linearly scales with the two observables. The coefficients $\alpha$ and $\beta$ determine the emphasis the tree places on each feature, with larger coefficients resulting in denser binning along the corresponding dimension.

\section{Results}\label{sec:results}

Fits were done for both the single power law (SPL) and the SPL+cutoff model to describe the measured astrophysical neutrino spectrum. The sensitive energy range for the measurements is determined through a reweighting method. Additionally, the piecewise neutrino flux is measured in five energy bins.  

\subsection{Single power law model}

\begin{figure*}
\includegraphics[width=0.45\textwidth]{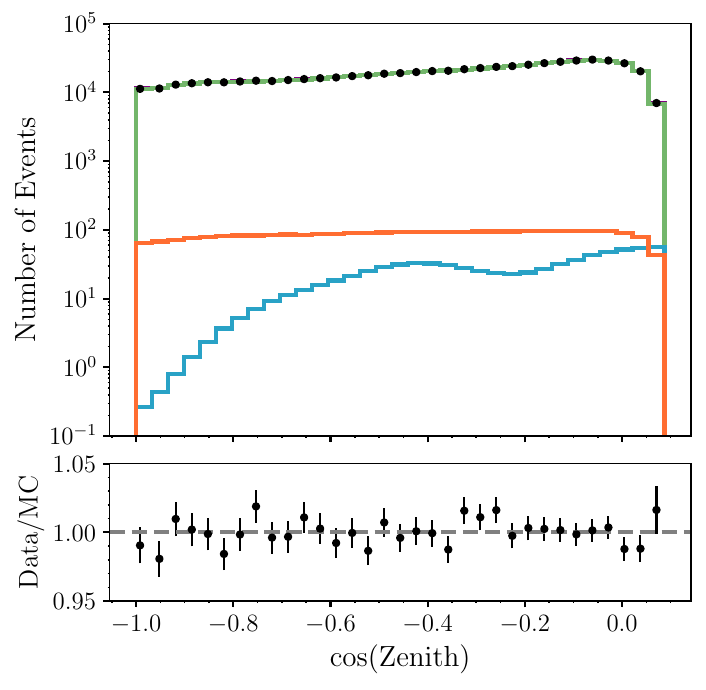}
\hfill
\includegraphics[width=0.46\textwidth]{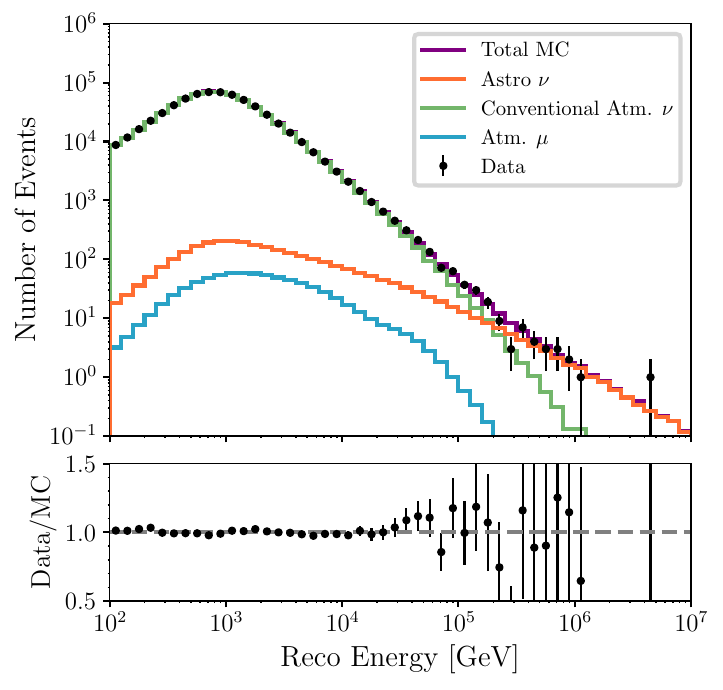}
    \caption{Comparison of observed and expected events distributions for data and simulations for the Northern Tracks sample, under the joint best-fit single power-law (SPL) model. Astrophysical neutrinos are indicated in orange, conventional atmospheric neutrinos in green, and atmospheric muons in blue. The prompt atmospheric neutrino flux is fitted to zero and is therefore not shown.}
    \label{fig:datamc_SPL_NT}
\end{figure*}

\begin{figure}
    \centering
    \includegraphics[width=0.99\linewidth]{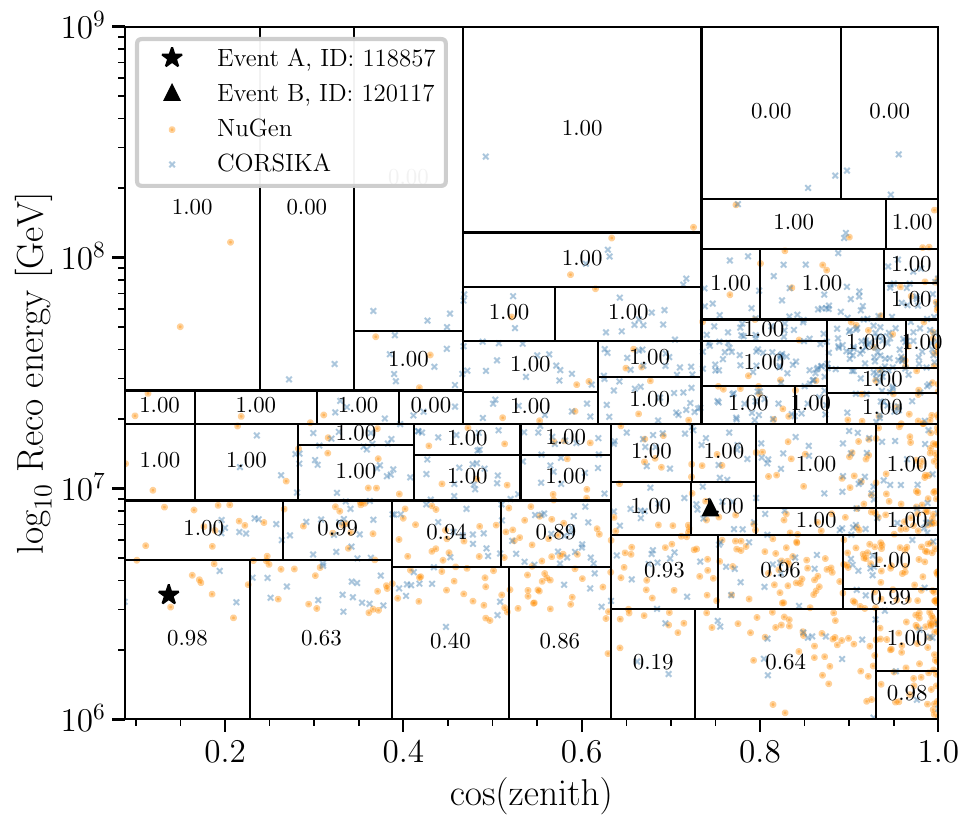}
    \caption{The non-uniform binning used for this event selection is shown by black grids obtained from a decision tree trained on simulations. The two observed events are shown as a black star and a triangle, respectively. Unweighted individual Monte Carlo events are displayed as orange (neutrino) and blue (atmospheric muon), and each bin is required to contain at least two unweighted events. The numbers in each bin are the signalness S/(S+B), computed using the results from the joint SPL fit.
    }
    \label{fig:nonuniform_binning_SPL}
\end{figure}

The SPL model assumes the following spectral shape:
\begin{equation}
\Phi(E_\nu) = \astroNorm \times \Big(\frac{E_\nu}{\SI{100}{TeV}}\Big)^{-\gammaAstro}
\end{equation}
where $\astroNorm$ (per-flavor) has units of $C_0 = 10^{-18} \times \si{GeV^{-1}. cm^{-2}. s^{-1}. sr^{-1}}$. A flavor ratio of $(\nu_e:\nu_\mu:\nu_\tau) = (1:1:1)$ is assumed at Earth~\cite{Theory:2008:review-of-high-energy-neutrino-astronomy-observe-flavor-ratio-120-to-111}. In the fit, all best-fit nuisance parameters stayed within their bounds, with small deviations from the nominal values, indicating a reliable fitting result. The best-fit prompt flux normalization is zero, consistent with the results from previous diffuse analyses~\cite{Diffuse:2021:7.5-year-HESE, Diffuse:2022:9.5-year-diffuse-numu, Diffuse:2020:6-year-cascades}. The best-fit downgoing background normalization is $\muonNormDPeV = 0.05^{+1.45}_{-0.05}$. The best-fit physics parameters for the SPL model are
\begin{align*}
    \astroNorm &= 1.65^{+0.28}_{-0.39} \\
    \gammaAstro &= 2.36^{+0.09}_.
\end{align*}
The Northern Tracks distributions assuming the best-fit SPL model are shown in Fig.~\ref{fig:datamc_SPL_NT}. The prompt flux is not visible as the normalization is 0. Data and MC show an overall good agreement, but a percent-level mismatch is present in the near-vertical low-energy bins around \SI{1}{TeV}. This issue has been investigated previously~\cite{Thesis:2021:Joeran-diffuse-numu}; the best-fit spectrum of astrophysical neutrinos is not affected by the mismatch. Here, the focus is on the spectral shape in the PeV region, so this data/MC mismatch at a few TeV has even less influence on the results. The statistical pull between data and best-fit MC expectation for all bins containing data is also studied. The distribution of $(\text{data}-\text{MC})/\sqrt{\text{MC}}$ is found to follow a Gaussian distribution of $\mathcal{N}(\mu=0.02, \sigma=1.03)$, indicating an overall reasonable data-MC agreement for most bins.

Visualizing the best-fit distributions for downgoing tracks is not easy due to the non-uniform binning. Instead of displaying each best-fit component in the 2D observable space, we show the signalness, $S/(S+B)$, for each bin in Fig.~\ref{fig:nonuniform_binning_SPL} based on the best-fit SPL flux, where $S$ and $B$ are the expected signal and background contributions per bin, respectively. 
Because the best-fit $\muonNormDPeV$ is small, (\ie, $0.05^{+1.45}_{-0.05}$), the overall signalness is relatively high. The number in each bin of Fig.~\ref{fig:nonuniform_binning_SPL} shows the signalness. Below \SI{5}{PeV}, the signalness is generally lower due to the presence of atmospheric muons, while it increases at higher energies. Due to the small number of MC events per bin, the associated signalness values have large statistical uncertainties, typically within 40\% below \SI{10}{PeV}, maximizing at 100\% for high-energy bins with only two events. The Monte Carlo event points are unweighted, while the signalness includes the weighting.   The two black markers in Fig.~\ref{fig:nonuniform_binning_SPL} show the locations of the two observed events in the signal region, denoted as Event A and Event B. Event A has a reconstructed energy $\etrunc=\SI{3.5}{PeV}$ and zenith $\cos\theta_{\text{zen}}=0.14$, and Event B has $\etrunc=\SI{8.1}{PeV}$ and $\cos\theta_{\text{zen}}=0.74$. Using the best-fit SPL flux and incorporating MC statistical uncertainties, Event A has a signalness of $0.975 \pm 0.34$, and Event B has a signalness of $1.0 \pm 0.47$, suggesting a possible astrophysical origin. The signalness is capped at 1.0 to ensure a physical interpretation.

\begin{figure}
    \centering
    \includegraphics[width=0.99\linewidth]{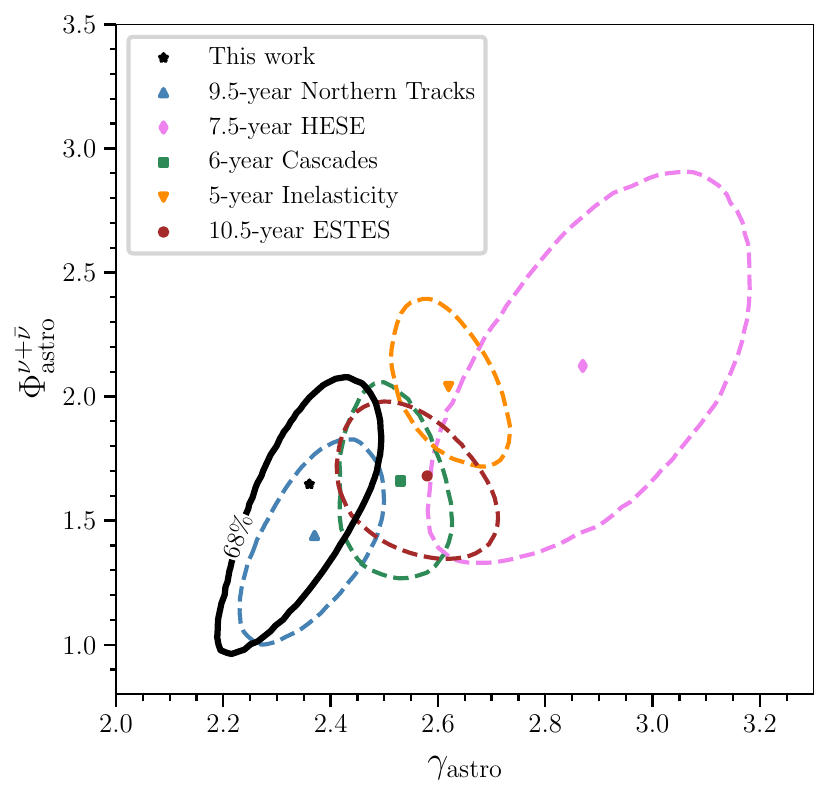}
    \caption{Best-fit single power-law results (solid markers) and profile likelihood landscapes (68$\%$ contours) as a function of spectral index and per-flavor normalization. This result is shown in black, together with previous results from 9.5-year Northern Tracks \cite{Diffuse:2022:9.5-year-diffuse-numu}, 7.5-year HESE \cite{Diffuse:2021:7.5-year-HESE}, 6-year Cascades \cite{Diffuse:2020:6-year-cascades}, 5-year inelasticity study \cite{Diffuse:2019:TeV-inelasticity-measurement}, and 10.5-year ESTES~\cite{Diffuse:2024:10-year-ESTES}. Our result is consistent with the previous Northern Track results.  The shift in best-fit values is driven by newer Monte Carlo simulations for Northern Tracks and the two observed high-energy downgoing events in our sample.}
    \label{fig:SPL_2Dscan}
\end{figure}

\begin{figure}
    \centering
    \includegraphics[width=0.99\linewidth]{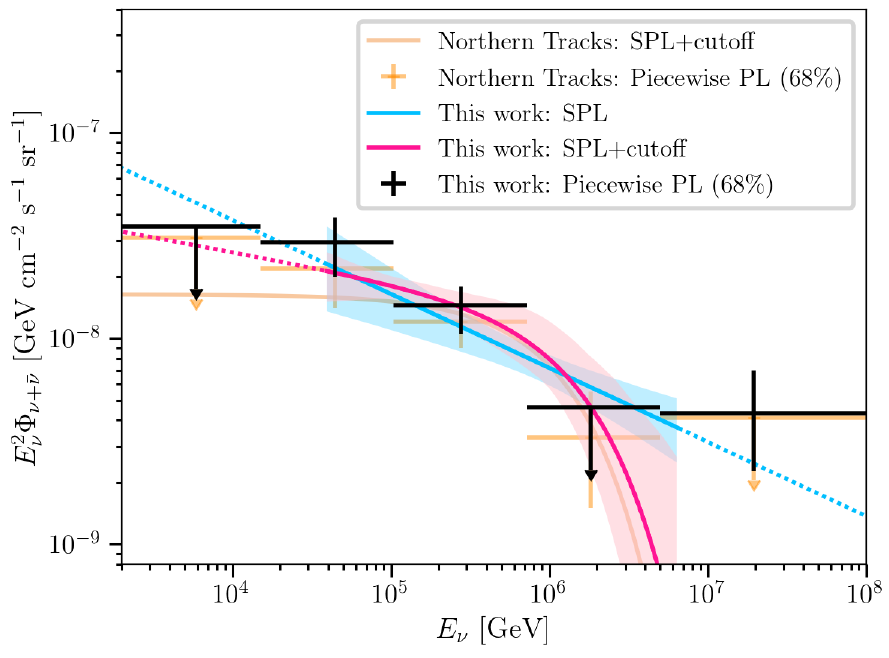}
    \caption{Best-fit per-flavor astrophysical neutrino flux as a function of neutrino energy under various model assumptions. The blue band shows the SPL results with $\pm 1\sigma$ uncertainties. The left (39 TeV) and right (6.3 PeV) boundaries of the band represent the sensitive energy range for this analysis under the SPL hypothesis. The best-fit SPL+cutoff flux and uncertainty are shown by the pink line and band. Outside the sensitive energy range, flux models are represented by dashed lines. The black crosses represent differential fluxes assuming an $E^{-2}$ spectrum for each bin. Uncertainties are at $68\%$ C.L. and upper limits are at $90\%$ C.L. The yellow crosses and lines represent measurements from the 9.5-year Northern Track paper \cite{Diffuse:2022:9.5-year-diffuse-numu}.}
    \label{fig:bestfit_fluxes}
\end{figure}

These SPL measurement results, $\astroNorm = 1.65^{+0.28}_{-0.39}$ and $\gammaAstro = 2.36^{+0.09}_{-0.10}$, are compatible with the previous 9.5-year Northern Tracks (NT) measurement: $\astroNorm^{\mathrm{NT}}=1.44^{+0.25}_{-0.26}$, $\gammaAstro^{\mathrm{NT}}=2.37^{+0.09}_{-0.09}$. The slightly higher normalization is mainly due to the removal of IC59 data and the updated $\nu_\tau$ treatment. A consistency check is performed by performing the SAY likelihood fitting using the updated Northern Tracks alone, resulting in a normalization of 1.70 and a spectral index of 2.42 with uncertainties similar to those in Ref. \cite{Diffuse:2022:9.5-year-diffuse-numu}. The two high-energy events lead to a slight hardening of the spectrum, but the change is not statistically significant.

The best-fit SPL parameters and the correlation between the spectral index and per-flavor normalization are shown in Fig.~\ref{fig:SPL_2Dscan}. The 1$\sigma$ contour, obtained using the profile likelihood technique, is shown as the black ellipse. A positive correlation is typically expected between $\astroNorm$ and $\gammaAstro$ for throughgoing tracks as a higher normalization compensates for the decrease of event counts due to a softer spectral index. Measurements from other previous or concurrent analyses are also included.  The different results are largely compatible with each other. Some differences between the different measurements could be because the various analyses are sensitive to different energy regions and flavors of neutrinos~\cite{Review:2022:nu-astronomy}. This work obtains a slightly larger contour than the Northern Tracks sample. The most likely reason is the low statistics in MC simulations in the downgoing sample, which widened the contour due to the use of the effective likelihood. The relatively large prior width (\ie, 1.40) of $\muonNormDPeV$ could also contribute to this widening.

The best-fit SPL model from this work is shown by the blue line in Fig.~\ref{fig:bestfit_fluxes}, with the blue error band representing the 68\% uncertainty. The width of the error band defines the sensitive energy range, while the spectrum outside this range is depicted by dashed lines. The sensitive energy range indicates where the astrophysical neutrino spectrum measurements are valid, calculated using the method from previous Northern Tracks analyses~\cite{Diffuse:2016:6-year-northern-track, Diffuse:2022:9.5-year-diffuse-numu} for consistency. This approach reweights astrophysical neutrino MC events by their contribution to the likelihood difference between a background-only fit and a full fit to the data. The sensitive energy range is defined as the energy range that captures the central 90\% of these reweighted events, which is [\SI{39}{TeV}, \SI{6.3}{PeV}] for this analysis. In comparison, the 9.5-year paper reports a sensitive range of [\SI{15}{TeV}, \SI{5}{PeV}]. The increase in the upper bound here is expected because of the additional sensitivity at high energies.  The change in the lower bound come mainly from  updates in Northern Tracks analysis (dropped IC59, new $\nu_\tau$ treatment, and new MCEq).

\subsection{Single power law with a cutoff}

Assuming the astrophysical spectrum could also be described by a SPL with an exponential cutoff, the following SPL+cutoff model (assuming a per-flavor normalization) is fitted to the data to measure the cutoff energy, given by $\cutoffEnergy$:
\begin{equation}
    \Phi(E_\nu) = \astroNorm \times \Big(\frac{E_\nu}{100 \mathrm{TeV}}\Big)^{-\gammaAstro} \times e^{-E_\nu/\cutoffEnergy}.
\end{equation}
The best-fit downgoing muon background normalization $\muonNormDPeV$ is $1.1^{+1.06}_{-0.69}$, larger than the SPL best-fit result. The best-fit physics parameters for the SPL+cutoff model are
\begin{align*}
    \astroNorm &= 1.90^{+0.37}_{-0.42} \\
    \gammaAstro &= 2.14^{+0.20}_{-0.22} \\
    \cutoffEnergy/\si{PeV} &= 1.83^{+6.92}_{-0.81}.
\end{align*}
The best-fit SPL+cutoff model is displayed in Fig.~\ref{fig:bestfit_fluxes} as the pink curve and pink uncertainty band. The same energy range is assumed as for the SPL fit. The result from the 9.5-year Northern Tracks analysis, which corresponds to $\cutoffEnergy^{\text{NT}}/\si{PeV} = 1.25^{+1.72}_{-0.56}$, is shown by the yellow curve. Our result prefers a slightly higher cutoff in comparison, albeit with a larger confidence interval. The presence of the two high-energy events shifts the measurements and confidence interval toward higher energies. The larger interval could be partly due to the use of the SAY likelihood and the relatively large prior width for $\muonNormDPeV$.

\begin{figure}
    \centering
    \includegraphics[width=0.99\linewidth]{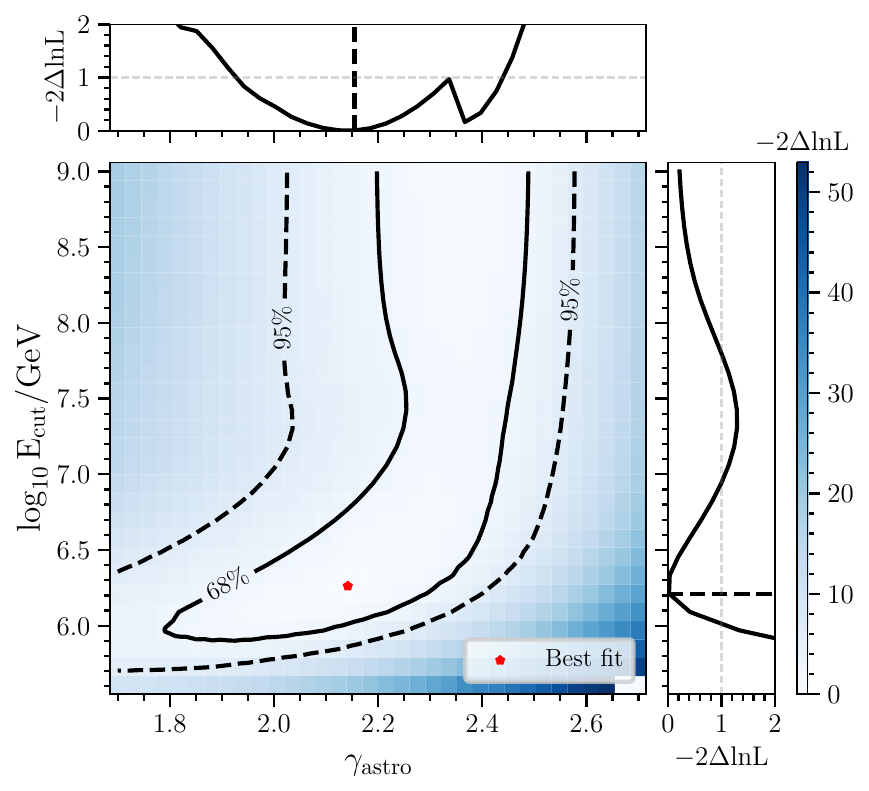}
    \caption{Profile likelihood landscape of cutoff energy vs. spectral index under the SPL+cutoff model. The red marker shows the best-fit result. Subpanels show corresponding 1D likelihood landscapes. The extension of the contours into the $E_{\mathrm{cutoff}}\rightarrow \infty$ region, together with the dip of spectral index at 2.36 in the 1D profile likelihood, indicate a similar preference for the SPL scenario. There is no statistical preference for a SPL+cutoff model.}
    \label{fig:SPLwCutoff_2Dscan}
\end{figure}

The correlation between $\cutoffEnergy$ and $\gammaAstro$ has been studied using a profile likelihood scan, as shown in Fig.~\ref{fig:SPLwCutoff_2Dscan}. The 2D likelihood contours are obtained using Wilks' theorem. The two sub-panels show $-2\Delta \ln \mathcal{L}$ values obtained by scanning over a grid of values for each parameter. Near the best-fit value (red marker), the 2D contour exhibits a positive correlation between the cutoff energy and the spectral index. This is expected because a harder $\gammaAstro$ leads to fewer events at higher energies, and the cutoff energy increases to compensate for the loss of events. A second minimum occurs when $\cutoffEnergy \rightarrow \infty$ and $\gammaAstro\approx 2.36$, which corresponds to the SPL best-fit results described before. This SPL scenario is less preferred compared to the SPL+cutoff best-fit results, where the SPL+cutoff finds an increase in $-2\Delta \ln \mathcal{L}$ by 0.15 (see Section~\ref{subsec:model-comparison}). 

Due to the relatively complex likelihood landscape, interpreting the confidence interval for the best-fit $\cutoffEnergy$ should be done carefully.  The cutoff energy is closely tied to the number of background cosmic-ray muons: the more background muons, the lower the cutoff energy. 


The $1\sigma$ upper bound at \SI{8.75}{PeV} is associated with the global minimum at $\cutoffEnergy=\SI{1.83}{PeV}$. The second, shallower minimum at infinity has an associated confidence interval of $[\SI{62}{PeV}, \infty]$. Therefore, the confidence interval for $\cutoffEnergy$ includes two disconnected regions. The interval of $[\SI{1.02}{PeV}, \SI{8.75}{PeV}]$ is reported as the final result because it contains the best-fit value.

We have examined the correlation between other parameters with the cutoff energy in the above 1D profile likelihood scan, which provides insight into the similar preference between the SPL and the SPL+cutoff models. When $\cutoffEnergy$ increases from \SI{1}{PeV} to \SI{1000}{PeV}, while most parameters remain constant, $\muonNormDPeV$ migrates from 1.1 to 0.05, with the latter being its best-fit value in the SPL model. The large prior width of $\muonNormDPeV$ (1.40) allows the muon background normalization parameter to almost freely scale the background distribution to accommodate the data under various flux assumptions. This flexibility leads to a similar preference between the SPL and the SPL+cutoff models, resulting in the two observed minima in the profile likelihood scan. In extreme cases, when $\muonNormDPeV$ has a high nominal value (e.g., $\gtrsim 2$) and a very small prior width, the observed two events can be fully accounted for by muon backgrounds, and the SPL model would be disfavored. This scenario would result in a single minimum, indicating a distinct preference for the SPL+cutoff model.

We examined the correlation between other parameters and the cutoff energy in the 1D profile likelihood scan, which provides insight into the similar preference between the SPL and SPL+cutoff models. 

\subsection{Piecewise flux}

\begin{figure}
    \centering
    \includegraphics[width=0.75\linewidth]{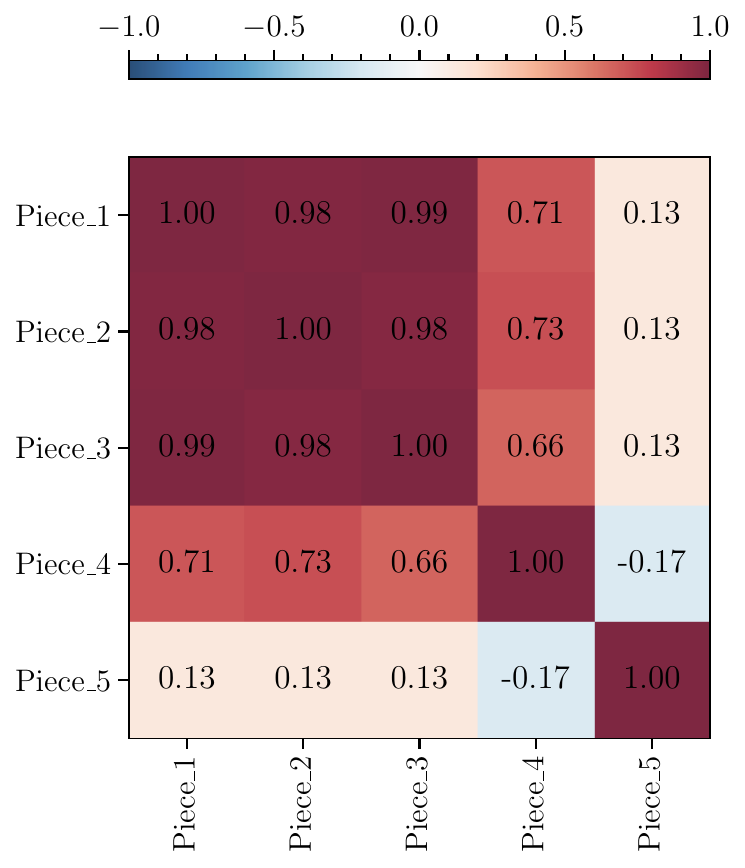}
    \caption{The correlation matrix between the different energy bins in the piecewise flux unfolding.}
    \label{fig:piecewise_corr}
\end{figure}

A more detailed spectrum can be found by using  piecewise flux unfolding to obtain a binned description of the astrophysical neutrino spectrum. The neutrino energy is divided into five bins, and it is assumed that the neutrino spectrum in each bin follows $E_\nu^{-2}$. The normalization $\phi_i$ (with a unit of $C_0$) for each bin $i$ is a free parameter. The bin edges used here are the same as those used in the Northern Tracks analysis.  Only five energy bins are used because throughgoing tracks have a limited energy resolution~\cite{Reco:2014:energy-reco-paper}. The total per-flavor astrophysical neutrino flux can be written as
\begin{equation}
    \Phi(E_\nu) = \sum\limits_{\text{bin i}}^{5} \chi(E_\nu)\cdot \phi_{i}\cdot  \times \left( \frac{\mathrm{E}_{\nu,i}}{\SI{100}{TeV}} \right)^{-2.0}
\end{equation}
where $\chi(E_\nu) = 1$ if  $E_{\rm low}^i < E_\nu < E_{\rm high}^i$ and $\chi(E_\nu) = 0$ otherwise. Here, $E_{\rm low}^i$ and $E_{\rm high}^i$ are the high and low bin edges respectively,
and $E_{\nu,i}$ is the neutrino energy in bin $i$. All normalizations $\phi_i$ are allowed to vary simultaneously. The best-fit results from this work and the 9.5-year Northern Tracks analysis are compared in Fig.\ref{fig:bestfit_fluxes}. For $\phi_1$ and $\phi_4$ in this work, which have best-fit values of 0, 90\% C.L. upper limits are determined. While this result is visually similar to the previous result, it is notable that the best-fit $\phi_4$ becomes zero and $\phi_5$ becomes non-zero. This difference is primarily due to the two observed events in the downgoing sample, along with the anticorrelation between $\phi_4$ and $\phi_5$ (correlation coefficient is $-0.17$, as shown in Fig.\ref{fig:piecewise_corr}). The two observed downgoing events have $\log_{10}\etrunc/\mathrm{GeV}$ at 6.5 and 6.9, respectively, corresponding to MC events in bin 5. Therefore, $\phi_5$ is preferred to be higher, driving $\phi_4$ to smaller values due to the anticorrelation.

\subsection{Model comparison}\label{subsec:model-comparison}

To determine which spectral model is preferred by the data, hypothesis testing is performed between the SPL ($H_0$) and the SPL+cutoff model ($H_1$). Given the limited data and MC statistics in the downgoing sample, pseudo-experiments are conducted to obtain accurate results instead of directly using Wilks' theorem. Specifically, 2500 toy datasets are sampled from the best-fit SPL flux assuming Poisson fluctuation in each analysis bin. For each toy dataset, the test-statistic $TS=-2\ln \left(\mathcal{L}_{H_0}/\mathcal{L}_{H_1}\right)$ is computed. Some likelihood fits during the pseudo-experiments failed because of the relatively flat SAY likelihood landscape around the minimum, where rounding errors make gradient calculations challenging. Due to the flat landscape, the likelihood values for these failed fits are still valid as the likelihood difference from the true minimum is small. Using successful pseudo-experiments (around 50\% of all trials) where likelihood fits succeed during both the SPL and SPL+cutoff measurements, the $\pval$ for the observed $TS$ is found to be 0.62. This result is similar to the $\pval$ computed using all pseudo-experiments (0.61) and is also similar to the $\pval$ computed assuming Wilks' theorem (0.7). Therefore,  this analysis fails to reject the SPL hypothesis, and there is no strong evidence to prefer an SPL+cutoff model.

\begin{table}
    \centering
    \renewcommand{\arraystretch}{1.5} 
    \begin{tabular}{c||ccc}
        \toprule
        Analysis & $\cutoffEnergy$ (\si{PeV}) & $\pval$ & Sensitive energy range \\
        \midrule
        This work & $1.83^{+6.92}_{-0.81}$ & 0.62 & \SI{39}{TeV} -- \SI{6.3}{PeV} \\
        NT~\cite{Diffuse:2022:9.5-year-diffuse-numu} & $1.25^{+1.72}_{-0.56}$ &  0.061 & \SI{15}{TeV} -- \SI{5}{PeV} \\
        HESE~\cite{Diffuse:2021:7.5-year-HESE} & $5.00^{+\infty}_{-3.8}$ & 0.71 & \SI{70}{TeV} -- \SI{1.9}{PeV} \\
        Cascades~\cite{Diffuse:2020:6-year-cascades} & $2.51^{+17.44}_{-1.51}$ & 0.34 & \SI{16}{TeV} -- \SI{2.6}{PeV} \\
        \bottomrule
    \end{tabular}
    \caption{Comparison of measurements of $\cutoffEnergy$ and $\pval$ for rejecting the SPL model compared to a SPL+cutoff model.}
    \label{table:previously-measured-cutoff-energy}
\end{table}

Measurements from different analyses for the cutoff energy and the corresponding $\pval$ between SPL \textit{vs.} SPL+cutoff is listed in Table~\ref{table:previously-measured-cutoff-energy}. All measurements of $\cutoffEnergy$ are compatible with each other within their uncertainties. 
The lower bound at \SI{1.02}{PeV} is relatively stringent. The Northern Tracks analysis sets the tightest constraint on the cutoff energy and observes the smallest $\pval$. Compared to the previous Northern Tracks results, this analysis obtains a larger $\pval$ mainly due to the two observed high-energy events with relatively high signalness, indicating a similar preference for the SPL model. Additionally, the complex SAY likelihood landscape and the relatively large prior width of $\muonNormDPeV$ decrease the power to differentiate the two models, which could contribute to a larger $\pval$. Furthermore, it will be shown later that one of the observed events could be of atmospheric origin, and a preference for SPL+cutoff could have been stronger.

\subsection{Event characteristics and point source searches}



\begin{table*}[t]
    \centering
    \renewcommand{\arraystretch}{1.5} 
    \setlength{\tabcolsep}{4pt} 
    \begin{tabular}{c||c|c|c|c|c|c|c|c|c}
        \toprule
        Name & Date & MJD & $\etrunc$ & $\theta_{\text{zen}}$ & $\dit$ & Stochasticity & $N_{\text{hits}}$ & RA [$^\circ$] & Dec [$^\circ$] \\
        \midrule
        Event A & Nov. 2011 & 55868.649 & \SI{3.47}{PeV} & $82.0^\circ$ & \SI{2036}{m} & 3.10 & 1 & $61.35^{+1.14}_{-1.28}$ & $-14.09^{+1.08}_{-1.4}$ \\
        Event B & May 2012 & 56058.124 & \SI{8.13}{PeV} & $42.3^\circ$ & \SI{1637}{m} & 3.08 & 1 & $115.01^{+0.41}_{-0.30}$ & $-48.14^{+0.28}_{-0.19}$ \\
        \bottomrule
    \end{tabular}
    \caption{Measured quantities for the two downgoing events in the signal region. $N_{\text{hits}}$ is the number of correlated IceTop hits. $\dit$ is the closest distance from the muon track to the center of IceTop, which is one of the variables in the IceTop inefficiency model. The event times are given both as calendar dates and in Modified Julian Date (MJD), along with the celestial coordinates (RA and Dec).}
    \label{table:observed-two-events}
\end{table*}

The characteristics of the two events are summarized in Table~\ref{table:observed-two-events}. The quoted uncertainties represent 90\% confidence intervals. Both events are also found in the newest versions of the ``PSTracks'' sample~\cite{PS:2023:extended-source-in-galactic-plane, PS:2024:nu-from-hard-xray-AGN, IceCube-HAWC:2024:joint-search-galactic-cr-accelerator}, although they were not identified in older versions of the sample~\cite{IceCube:2021xar}. In the new versions, the application of an IceTop veto and stochasticity cuts allow them to be singled out as likely signal.

Event A is near the horizon with a large $\dit$ (closest distance from the muon track to the center of IceTop), while Event B is more vertical, with a higher energy. The characterization of Event A turns out to be challenging. First, the associated IceTop activity shows three hits along the direction of the track, but they are found to be slightly outside the veto time window, with $\Delta t$ around $\SI{2400}{ns}$. Because the veto time window is fixed in this analysis, possible additional delays in the arrival times of IceTop hits at larger zenith angles or larger $\dit$ due to shower curvature are not taken into account. Therefore, the visually correlated IceTop hits for Event A suggest that this event could have an atmospheric origin.

Another complication arises during the direction reconstruction for Event A. The position of the observed signal is close to the dust layer, the region with the strongest absorption~\cite{Det:2013:measure-ice-property-SPICEMie}. Although the default SplineMPE reconstruction gives a zenith angle of $\theta=82^\circ$, different reconstruction algorithms, including the likelihood-based sky scan used during the point source searches, give a zenith from 5 to 10 degrees lower~\cite{Reco:2004:muon-track-reco-AMANDA, Reco:2004:muon-track-reco-AMANDA, PS:2023:IceCat-1, Reco:2013:DirectFit}. 

In contrast, different reconstructions for Event B generally agree with each other within the uncertainty. Additionally, its $dE/dx$ values include three bins exceeding \SI{2000}{GeV/m}, indicating a more robust estimation of stochasticity (less likely due to artifacts during reconstruction) compared to Event A, which only has one bin with large energy losses. If Event B is an astrophysical neutrino, the neutrino energy would need to exceed \SI{8}{PeV}, making it one of the highest energy neutrino events seen by IceCube.

With these caveats, this section describes the searches for astrophysical objects spatially coincident with Event A and Event B. 

Since 2016, IceCube has been issuing real-time alerts for high-quality tracks with signalness >30\% (assuming an $E^{-2.19}$ astrophysical neutrino spectrum) to follow-up observatories across multiple wavelengths~\cite{PS:2017:realtime-alert-system, PS:2020:Next-gen-realtime-alerts}. IceCat-1~\cite{PS:2023:IceCat-1} is the first catalog of historical IceCube alerts with 275 events (2011-2020) processed using the realtime event selections. It reported uncertainty contours from a likelihood-based sky scan (``Millipede scan'') and conducted point source searches using several high-energy astrophysical catalogs. In this paper, a point source check is performed using the same catalogs and Millipede scan for consistency. The catalogs include four high-energy $X$-ray and $\gamma$-ray surveys: 4FGL-DR4~\cite{Catalog:2023:4FGL-DR4}, 3HWC~\cite{Catalog:2020:3HWC}, Swift-BAT~\cite{Catalog:2018:Swift-BAT}, and TeVCat~\cite{Catalog:2008-TeVCat}; they may have overlapping sources.

The point source check here is qualitative; no $\pval$ is computed, and no temporal correlations are studied. It compares the locations of astrophysical sources from the above catalogs with the directional uncertainty contours for each event. Fig.~\ref{fig:event_scan} shows the results. Event A (left) has one associated source within the 90\% C.L. contour, reported in both the 4FGL and Swift catalogs, with PKS 0403-13 as the counterpart.
Event B (right) has no associated sources. The Millipede contours for Event A (4.55 square degrees at 90\%) are much larger than those for Event B (0.26 square degrees at 90\%) due to reconstruction uncertainties. Therefore, the association of Event A with the BL Lac object could be coincidental and should be interpreted cautiously.

For comparison, the IceCat-1 study found no associated sources for 139 out of 275 neutrino events (a fraction of 50.5\%).
Observing a source within the uncertainty contour does not necessarily imply the neutrino is produced from the associated source; the IceCat-1 study found that the fraction of neutrino alerts with associated objects is consistent with random coincidence. 

In summary, although Event A has a possible associated object within the 90\% uncertainty contour, it is also possible that neither Event A nor B is correlated with any astrophysical sources from the four catalogs. The lack of correlation could suggest that the environment near the astrophysical counterpart is optically thick to $X$-rays and $\gamma$-rays. The source could also be far away (e.g., $z > 1$), where photons are absorbed by the interstellar medium. Generally, a large fraction of high-energy diffuse neutrinos observed at IceCube have no known sources, except those associated with the blazar TXS 0506+056~\cite{PS:2018:TXS0506-056-blazar} and the active galaxy NGC 1068~\cite{PS:2022:neutrino-from-NGC-1068}. Stringent upper limits on luminous sources like blazars~\cite{PS:2022:blazar-contribution-to-diffuse-flux-upperlimit} and GRBs~\cite{PS:2022:GRB-limit-prompt-and-afterglow} suggest that low-luminosity objects with large number densities, such as low-luminosity AGNs and starburst galaxies, could be the dominant neutrino sources~\cite{Gen2:2023:icrc2023-overview, Review:2022:nu-astronomy}. 

\section{Conclusions}\label{sec:conclusions}

This work studied the astrophysical neutrino spectrum in the PeV region by obtaining a sample of downgoing high-energy throughgoing tracks from the Southern Sky. This work is the first analysis with IceCube that combines the use of the stochasticity variable and the IceTop surface veto to reject the large number of atmospheric muon background events, facilitated by a data-driven muon background estimation method. Using 9 years of data, the final sample contains two high-energy events with a relatively high signal-to-background ratio.
A joint effective likelihood fit is performed with the previous 9.5-year Northern Tracks sample~\cite{Diffuse:2022:9.5-year-diffuse-numu}.

Inside the sensitive energy range from \SI{39}{TeV} to \SI{6.3}{PeV}, the SPL measurements of $\astroNorm = 1.65^{+0.28}_{-0.39}$ and $\gammaAstro = 2.36^{+0.09}_{-0.10}$ are consistent with the Northern Tracks results, while having a larger normalization and a harder spectrum. The measured cutoff energy, $\cutoffEnergy = 1.83^{+6.92}_{-0.81}\ \si{PeV}$, is slightly higher than the previous results and has a wider confidence interval. A likelihood ratio test between the SPL \textit{vs.} SPL+cutoff models gives a $\pval$ of 0.62, indicating the SPL+cutoff model is not significantly preferred over a SPL. A model-free piecewise flux unfolding was performed in five neutrino energy bins, where a larger flux is observed at highest energies beyond \SI{5}{PeV} compared to the Northern Tracks results. 

Point source searches were performed in the direction of the two observed events using four X-ray and $\gamma$-ray source catalogs. One event, which has a relatively large directional uncertainty contour, is near the BL Lac object PKS 0403-13. This spatial correlation alone does not necessarily imply that the event originates from this source. The other event has no associated sources; this non-association is consistent with the assumption that a large fraction of the diffuse flux could originate from low-luminosity sources at large distances.


In the future, the stochasticity calculation, surface veto, and muon background estimation methods developed in this paper can be effectively applied to the next-generation IceCube-Gen2 detector~\cite{Gen2:2021:physics-and-design}. IceCube-Gen2 will significantly improve sensitivity to the PeV spectral cutoff with its larger fiducial volume ($\sim \SI{8}{km^3}$) and the additional radio antennas and scintillators as the surface veto~\cite{Gen2:2023:surface-array}. 


\begin{acknowledgements}
The authors gratefully acknowledge the support from the following agencies and institutions:
USA {\textendash} U.S. National Science Foundation-Office of Polar Programs,
U.S. National Science Foundation-Physics Division,
U.S. National Science Foundation-EPSCoR,
U.S. National Science Foundation-Office of Advanced Cyberinfrastructure,
Wisconsin Alumni Research Foundation,
Center for High Throughput Computing (CHTC) at the University of Wisconsin{\textendash}Madison,
Open Science Grid (OSG),
Partnership to Advance Throughput Computing (PATh),
Advanced Cyberinfrastructure Coordination Ecosystem: Services {\&} Support (ACCESS),
Frontera and Ranch computing project at the Texas Advanced Computing Center,
U.S. Department of Energy-National Energy Research Scientific Computing Center,
Particle astrophysics research computing center at the University of Maryland,
Institute for Cyber-Enabled Research at Michigan State University,
Astroparticle physics computational facility at Marquette University,
NVIDIA Corporation,
and Google Cloud Platform;
Belgium {\textendash} Funds for Scientific Research (FRS-FNRS and FWO),
FWO Odysseus and Big Science programmes,
and Belgian Federal Science Policy Office (Belspo);
Germany {\textendash} Bundesministerium f{\"u}r Bildung und Forschung (BMBF),
Deutsche Forschungsgemeinschaft (DFG),
Helmholtz Alliance for Astroparticle Physics (HAP),
Initiative and Networking Fund of the Helmholtz Association,
Deutsches Elektronen Synchrotron (DESY),
and High Performance Computing cluster of the RWTH Aachen;
Sweden {\textendash} Swedish Research Council,
Swedish Polar Research Secretariat,
Swedish National Infrastructure for Computing (SNIC),
and Knut and Alice Wallenberg Foundation;
European Union {\textendash} EGI Advanced Computing for research;
Australia {\textendash} Australian Research Council;
Canada {\textendash} Natural Sciences and Engineering Research Council of Canada,
Calcul Qu{\'e}bec, Compute Ontario, Canada Foundation for Innovation, WestGrid, and Digital Research Alliance of Canada;
Denmark {\textendash} Villum Fonden, Carlsberg Foundation, and European Commission;
New Zealand {\textendash} Marsden Fund;
Japan {\textendash} Japan Society for Promotion of Science (JSPS)
and Institute for Global Prominent Research (IGPR) of Chiba University;
Korea {\textendash} National Research Foundation of Korea (NRF);
Switzerland {\textendash} Swiss National Science Foundation (SNSF).
\end{acknowledgements}


%

\clearpage

\appendix

\section{IceTop Veto inefficiency}
\label{app:icetop}

The IceTop veto is only effective in regions where the expected IceTop signal is significant.  Several orders of magnitude in background rejection are required.  The effective region depends on several variables: the primary energy and composition of the cosmic-ray particle, the inclination (\ie, $\cos\theta$), the distance from the muon track to the center of IceTop ($\dit$), and the distance from the track to the closest IceTop tanks ($d_{\mathrm{closest}}$)~\cite{Diffuse:2011:first-icetop-veto-idea, Det:2013:IceTop-paper}.

For practical reasons, the effectiveness of the IceTop veto is quantified as ``inefficiency'': the fraction of atmospheric muons that pass the veto divided by the total number of events. A smaller inefficiency indicates a more effective veto. In this analysis, we parameterize the inefficiency using $\etrunc$ and $\dit$.  This inefficiency model is then fit to data. 

To model the inefficiency, an anti-stochasticity cut, $\stoch < 0.8$, is applied to the full data to remove any neutrino signals. This threshold is arbitrary and is later incorporated into the systematics uncertainty. The ranges and binnings for the modeling procedure are defined: $5.5 \leq \log_{10}\etrunc/\si{GeV} \leq 6.5$ with a bin width of $\Delta E = 0.025$, and $\SI{0}{m} \leq \dit \leq \SI{2500}{m}$ with a bin width of \SI{50}{m}. The IceTop inefficiency of the data for each bin $(i, j)$ is then calculated as $\ineff(d_i, E_j) = N_{\mathrm{pass}}(d_i, E_j)/N_{\mathrm{total}}(d_i, E_j)$. For each energy slice, the inefficiency model as a function of $\dit$ takes the form
\begin{equation}
    \ineff(\dit) = \frac{a(E)}{\left(\frac{\dit}{\SI{2000}{m}}\right)^{0.2}} \exp\left(-\frac{b(E)}{\left(\frac{\dit}{\SI{2000}{m}}\right)^{3.7}}\right)
    \label{eq:ineff-model}
\end{equation}
This form was chosen because it fits the data well, and has the limit $\ineff = 0$ when $\dit = 0$, motivated by the fact that events in all energy bins are fully vetoed within $\dit < \SI{500}{m}$. At larger distances, the inefficiency increases and then reaches a plateau, with height and location set by $a(E)$ and $b(E)$, respectively. An exponent of 3.7 maximizes the overall agreement between the model and the data. During the fit, for each energy slice $i$, the inefficiency model is first fit to data as a function of $\dit$, yielding a list of the best-fit parameters $a(E_i)$ and $b(E_i)$, which are then fit by a second-order polynomial as a function of $\log_{10}\etrunc$. To mitigate statistical uncertainties, the trends of $a$ and $b$ are required to be monotonic as energy increases, because the overall inefficiency should decrease with an increasing energy. This is achieved by imposing constraints during the fit: $a(E_{i+1}) \leq a(E_i)$ and $b(E_{i+1}) \geq b(E_i)$. Lastly, a floor of $10^{-5}$ is set for the inefficiency values throughout the energy-distance space. 

At the South Pole, snow accumulates on the surface of the ice at an average rate of \SI{20}{cm} per year, primarily due to drifting~\cite{Det:2013:IceTop-paper}. This reduces the sensitivity of IceTop tanks to the electromagnetic components of air showers without affecting the muon spectra. Consequently, there are fewer correlated IceTop hits in data from later years, shifting the inefficiency contours towards the upper-left direction. To account for the snow effect, different inefficiency models corresponding to each year's data are constructed. 

\section{Muon backgrounds}
\label{app:muons}

Estimating the muon background is challenging because of the limited statistics of simulations of combined in-ice detector and IceTop.  The combination of high cosmic-ray energies and low individual particle thresholds is computationally demanding. Relying solely on simulations also involves issues of unaccounted-for rare background events and data-MC mismatches. The larger challenge is due to the composition dependence of the veto and stochasticity.

At lower energies (\eg, $E_{\mu} < \SI{10}{TeV}$ at the surface), the muon bundle energy is typically shared by all muons within a bundle. At higher energies, leading muons, where a single muon carries most of the bundle energy, begin to dominate in bundles~\cite{CR:2016:characterize-atm-muon-flux-in-IC}. Leading muons typically come from rapid decays of energetic pions, kaons, or charmed particles in the early stages of shower development. The presence of leading muons increases the measured stochasticity because the energy losses from other non-leading muons are much smaller. For a given primary energy, protons are more likely to produce leading muons than iron, due to their higher energy per nucleon. Combined with the fact that protons create fewer muons on average and have a larger depth at shower maximum compared to irons, this leads to a correlation between stochasticity and the IceTop veto: muons from proton showers are generally more stochastic and leave fewer traces in IceTop.

\subsection{Procedure}\label{subsec:procedure}

These limitations motivate a data-driven atmospheric muon background estimation method. The idea is to construct a model for the passing probability of atmospheric muons as a function of stochasticity. This model is then fit to the passing fraction ($N_{pass}/N_{total}$) of data in the fitting region at low stochasticity. Before the modeling, two cuts are first applied to both data and simulations: $\ineff < 0.01$ and phase space $<$ 0.01. IceTop veto is not applied except when calculating the passing fraction of data. 

The CORSIKA-in-ice (no IceTop information) simulation~\cite{Simulation:1998:CORSIKA} is split into two populations, Single and Bundle, depending on $\fleading=E_{\mathrm{leading}}/E_{\mathrm{bundle}}$. Events in the Single population, which have a higher leading muon energy fraction, should pass the IceTop veto more easily because of a smaller IceTop response on average. These events should also have higher measured stochasticity as it is positively correlated with $\fleading$. Here, we refer to $\fleading$ as the ``singleness'' parameter, because an event with a higher singleness has an energy loss profile more similar to that of a single muon. The nominal singleness threshold is chosen to be 0.7. Different singleness thresholds lead to slightly different background estimates; this is treated as a systematic uncertainty.

\begin{figure}
    \centering
    \includegraphics[width=0.99\linewidth]{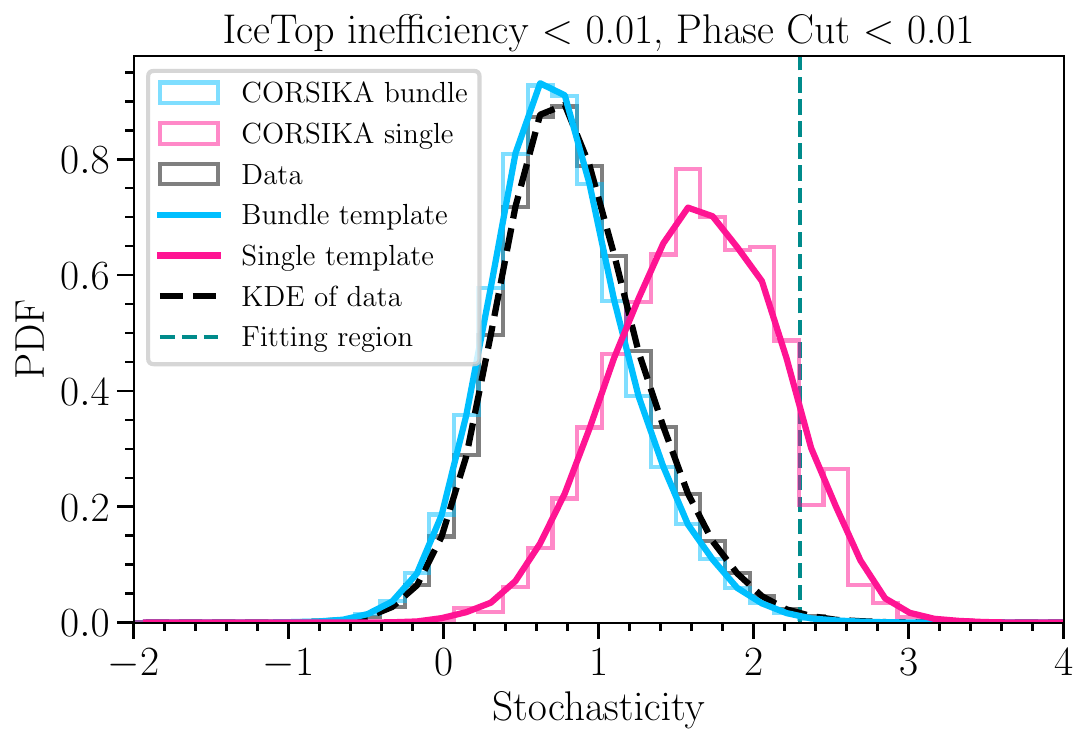}
    \caption{Stochasticity templates used in estimating the muon background. Cuts of Inefficiency $<$ 0.01 and Phase $<$ 0.01 are applied to both the data and CORSIKA events. Histograms representing various components are normalized, and Kernel Density Estimations (KDEs) are then performed on each distribution. The pink and blue curves represent the Single and Bundle templates, respectively. The KDE of data is shown in black. The region to the left of the vertical green line is the fitting region.}
    \label{fig:muon_templates}
\end{figure}

The stochasticity distributions for Single and Bundle atmospheric muons are shown in Fig.~\ref{fig:muon_templates}, containing 40 bins from -2 to 4.2. The same stochasticity binning is used when fitting for the passing probability model described later. To reduce statistical fluctuations, Kernel Density Estimations (KDEs) are performed on these distributions. The results are two stochasticity templates, referred to as $\KDEs(s)$ and $\KDEb(s)$, where $s$ is the stochasticity variable. Each template represents a normalized distribution of the corresponding muon population. The Bundle events have a much lower average stochasticity then Single events. 

The mixture of the two templates are then fit to the KDE of data (black-dashed curve in Fig.~\ref{fig:muon_templates}) to obtain the template normalizations. These normalizations characterize the relative contribution of each muon population in the data. The fit is performed in the region  stochasticity $<$ 2.3 (the fitting region), shown by the green line in Fig.~\ref{fig:muon_templates}.  This avoids the signal region, reducing the potential bias due to the possible presence of signal. The function $L(\yb)$ is minimized:
\begin{align}
    L(\yb) &= \sum\limits_{\mathrm{bins\ i}}\big(\KDEdata(s_i) - \KDEtot(s_i)\big)^2 \\
    \KDEtot(s_i) &= \yb \cdot \KDEb(s_i) \notag \\
    &\quad + (1-\yb) \cdot \KDEs(s_i) \label{eq:KDEtot}
\end{align}
where $s_i$ denotes the stochasticity for each bin, and $\KDEtot$ is the mixture of the two templates. The normalization for the Bundle template, $\yb$, is  the fraction of muon bundles in data, with 
$0 \ge \yb \le 1.$.  Because events can only be singles or bundles, the normalization of the Single template is therefore $1-\yb$. The best-fit Bundle normalization is $\yhatb = 0.927$: most of the data is  muon bundles. Different objective functions such as Poisson likelihoods lead to similar best-fit parameters.

\begin{figure}
    \centering
    \includegraphics[width=0.99\linewidth]{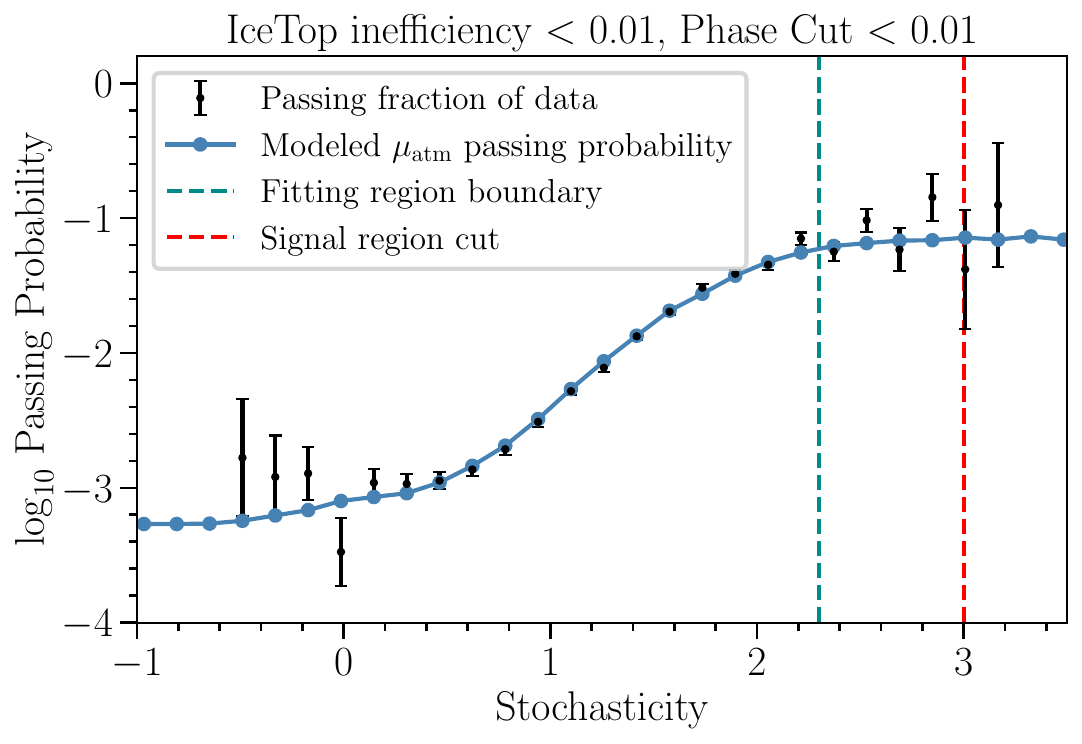}
    \caption{The best-fit passing probability of atmospheric muons as a function of stochasticity. Black markers represent the passing fractions of data in each bin, and the blue curve is the best-fit passing probability model (PPM). For the region where stochasticity > 2.5, a linear fit (not shown) is applied to further smooth the passing probability model. The green vertical line shows the model fitting region, and the red line shows the signal region boundary.}
    \label{fig:modeled_passing_proba}
\end{figure}

We then assign the Single and Bundle muon populations two constant probabilities of passing the veto, $\Ps$ and $\Pb$, respectively. They are free parameters to be determined. The probability for an atmospheric muon to pass the veto at each stochasticity bin is given by the Passing Probability Model (PPM), $\mathcal{P_\mathrm{pass}}$:
\begin{multline}
    \mathcal{P_\mathrm{pass}} (s_i;\Pb, \Ps) = \frac{\yhatb \cdot \KDEb(s_i)}{\KDEtot(s_i)} \cdot \Pb \\
    + \frac{(1 - \yhatb) \cdot \KDEs(s_i)}{\KDEtot(s_i)} \cdot \Ps
    \label{eq:passmodel}
\end{multline}
This PPM is fit to the passing fractions of data in the fitting region, shown by Fig.~\ref{fig:modeled_passing_proba}. The passing fractions of data assume binomial uncertainties. The fit aims to determine the values for $\Pb$ and $\Ps$ by minimizing the negative-log of the binomial likelihood $L$:
\begin{equation}
    L(\Pb, \Ps) = \prod\limits_{\text{bins i}} \binom{n_i}{k_i} \passmodel_i^{k_i} \big(1 - \passmodel_i \big)^{n_i-k_i} 
\end{equation}
where $n_i$, $k_i$ are the number of data before and after the veto in each stochasticity bin $i$, respectively. $\passmodel_i \equiv \mathcal{P_\mathrm{pass}} (s_i; \Pb, \Ps)$ denotes the value of the PPM in the given bin, bounded by
$0 \ge \Pb,\Ps \le 1.$.  The best-fit parameters are $\Phatb=0.0005$, $\Phats=0.0806$. After plugging them into Eq.~\ref{eq:passmodel}, the best-fit PPM for unblinded data is shown by the blue curve in Fig.~\ref{fig:modeled_passing_proba}. The residuals are near 0 with a reduced chi-square value of 1.36. The residuals are also symmetrically distributed around 0.

Although not visible, the best-fit PPM exhibits some dips in the region of stochasticity > 3.7, likely due to the oscillatory behavior of KDE estimations in regions with very few events. To address this, a linear fit is performed on the PPM in the stochasticity range from 2.5 to 3.5. The PPM then adopts the value from the linear fit result at stochasticity 3.0 and above. This final form of the PPM is used throughout the analysis, and is referred to as the nominal passing probability model $\mathcal{P}_{\mathrm{nominal}}$.

\subsection{Estimating the background}\label{subsec:estimated-background}

\begin{figure}
    \centering
    \includegraphics[width=0.99\linewidth]{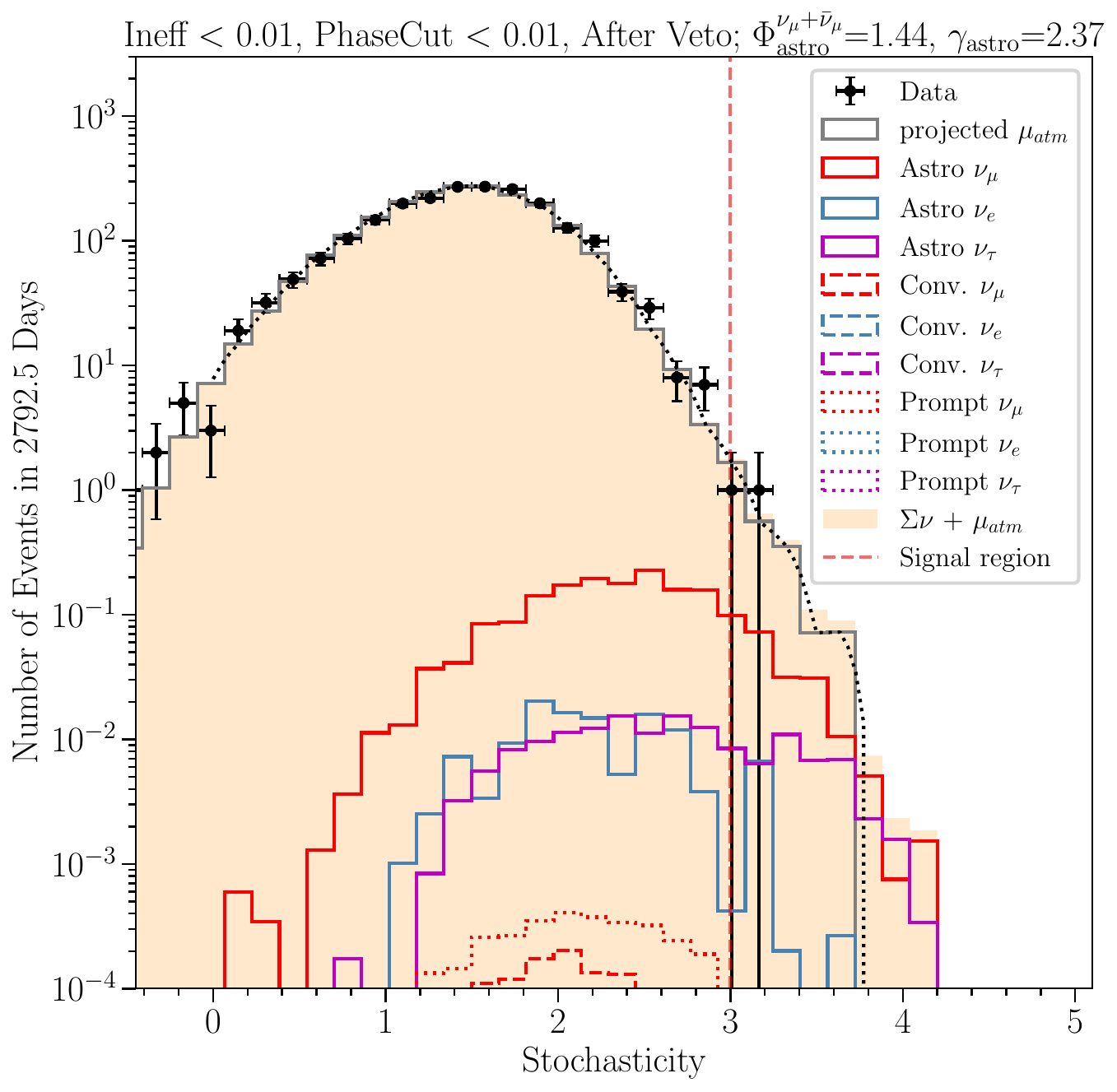}
    \caption{The stochasticity distribution for different components after the inefficiency cut, phase space cut, and IceTop veto are applied. The gray histogram shows the estimated atmospheric muon background for each stochasticity bin, and the dotted curve represents its cubic interpolation along bin centers. Inside the signal region (stochasticity $>$ 3.0), two events (black dots) are observed.  The integral below the dotted curve represents the nominal background estimation (1.97 events).}
    \label{fig:stochasticity_unblinded}
\end{figure}

After obtaining the nominal PPM, the number of atmospheric muon background events in the signal region can be calculated. Figure~\ref{fig:stochasticity_unblinded} shows the stochasticity distribution of data and neutrino MC after the inefficiency and phase space cuts, where the neutrino spectrum assumes the 9.5-year Northern Tracks best-fit results~\cite{Diffuse:2022:9.5-year-diffuse-numu}. Atmospheric neutrinos, although neglected in this analysis, as still shown for comparison. The red vertical line at 3.0 is the stochasticity cut. The black markers represent data that pass the veto. Therefore, the two data points to the right of the stochasticity cut are in the signal region. The atmospheric muon background in each stochasticity bin ($s_i$) is computed as follows:
\begin{equation}
    N_{\mathrm{BG}}(s_i)=N^{\mathrm{data}}_{\text{before veto}}(s_i) \cdot \mathcal{P}_{\mathrm{nominal}} (s_i)
\end{equation}
Because data before the veto potentially contain astrophysical signals, this is a conservative estimation of the number of background events. The calculated background events are shown as the gray histogram in Fig.~\ref{fig:stochasticity_unblinded}. To calculate the background in the signal region, a cubic interpolation over the midpoints of the histograms is performed (grey dotted curve), and the area under the curve above the stochasticity of 3.0 is then integrated. The number of background events in the signal region is 1.97, consistent with the two observed events. This background estimate is treated as a nuisance parameter with a Gaussian prior, having a mean of 1.97 and a width to be estimated in Section~\ref{subsec:muon-bg-estimation-uncertainties}.

The statistical uncertainties on this background estimate can be estimated by propagating the statistical uncertainties of model parameters $\yhatb$, $\Phatb$, and $\Phats$ into it. The parameter uncertainties can be estimated using Bayesian methods, \ie, Markov Chain Monte Carlo (MCMC)~\cite{Reco:2019:MCMC-intro}. Because the fit is performed first in $\yb$ and then in $\Pb$, $\Ps$ simultaneously, the uncertainty in $\yb$ is first estimated. The sampling distribution of $\yhatb$ is obtained by bootstrapping, which approximates a Gaussian $\mathcal{N}(0.927, \sigma=0.013)$. We then sample $\yb$ from this distribution and perform MCMC on $\Pb$ and $\Ps$ simultaneously conditioned on each sampled $\yb$:
\begin{align}
     \yb & \sim \mathcal{N}(\yhatb, \sigma=0.013) \\
     (\Pb, \Ps) & \sim \text{Posterior}(\Pb, \Ps | \yb)
\end{align}
From the posterior distributions, we obtain $\Pb = 0.0005 \pm 0.0001$, and $\Ps=0.0806 \pm 0.0023$. Propagating these uncertainties to the background estimation result, it follows a Gaussian $\mathcal{N}(1.97, \sigma=0.05)$. This uncertainty will be incorporated into the estimation of the prior width for the background normalization.

In addition to the normalization, the background distributions in the observable space (energy and zenith) are needed for subsequent likelihood fits for the neutrino spectrum. For this, the weight of each CORSIKA event is multiplied by its corresponding IceTop inefficiency. Using the nominal inefficiency model leads to discrepancies in the energy and zenith distributions between the inefficiency-weighted CORSIKA and data that pass the veto. The discrepancy arises from the correlation between the true IceTop inefficiency and the zenith angle. To account for this, a three-dimensional inefficiency model is constructed by performing a spline interpolation on the inefficiency of the full dataset as a function of $\etrunc$, $\dit$, and $\cos\theta$. An anti-stochasticity cut (stochasticity is less than 2.5) is applied to ensure the modeling is performed in the background-dominated region, although the choice of the cut value is found to have negligible effect on the modeling results. This 3D inefficiency model is only used to obtain the shape of the CORSIKA distribution in the observable space after the veto.

\section{Muon background uncertainties}
\label{app:muonuncertainties}

The muon background estimation method accounts for various systematic uncertainties. They are only propagated to the calculated background normalization, not to the shape of background distributions in observable space. Detector systematics for CORSIKA simulations are also excluded. Despite these limitations, systematics are subdominant for the estimated background, as the nominal background (1.97) has a much larger statistical uncertainty.
Only two events are observed in the signal region, again suggesting that statistical uncertainties dominate for this analysis.

\begin{figure}
    \centering
    \includegraphics[width=0.99\linewidth]{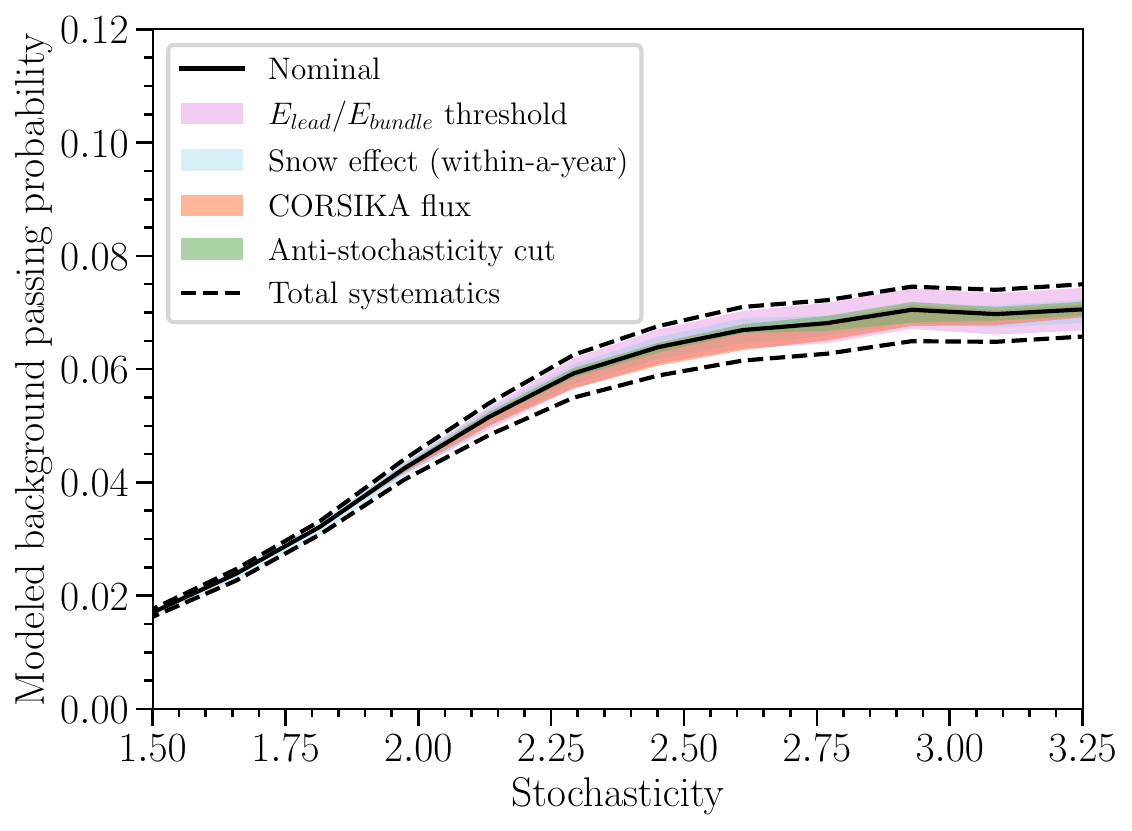}
    \caption{Systematic uncertainties affecting the best-fit passing probability curve. The uncertainties arise from factors including the threshold used to construct CORSIKA templates, snow accumulation on IceTop, the choice of the CORSIKA flux model, and the threshold of the anti-stochasticity cut when modeling the IceTop inefficiency. The solid black curve represents the nominal best-fit passing probability, while the dashed curves illustrate the total systematic uncertainty.}
    \label{fig:passing_proba_systematics}
\end{figure}

The singleness threshold for splitting CORSIKA simulations into two templates is set to 0.7 in the nominal case. We remodeled the PPM using singleness thresholds of 0.65 and 0.75, respectively. This variation is illustrated by the pink uncertainty band in Fig.~\ref{fig:passing_proba_systematics}. The PPM shows higher values in the high-stochasticity region for an increased singleness threshold due to the shift of the Single template. At higher singleness thresholds, the Single template shifts to higher stochasticity, leading to a higher $\yhatb$ because the Bundle template more closely resembles the data distribution. Consequently, the parameter $\Ps$ is fit to higher values to match the passing fraction of data in the fitting region (stochasticity $<$ 2.3). Therefore, the best-fit passing probability at high stochasticity, driven by the $\Ps$ value, increases.

Similar to the singleness threshold, setting the anti-stochasticity cut at 0.8 is also an arbitrary choice, so various alternative cut values are studied. We remodeled the PPM using anti-stochasticity cuts from 0.6 to 1.0, applied to both MC simulations and data. For each anti-stochasticity cut value, given that the inefficiency contours for each year's data differ from the nominal case, time-dependent inefficiency contours and cuts are constructed and applied, as outlined in Section~\ref{subsec:icetop-veto-and-inefficiency}. The results of the remodeling are shown by the green band in Fig.~\ref{fig:passing_proba_systematics}.

The effect of changing the CORSIKA primary flux assumption is studied. Two other fluxes, H4a~\cite{Model:2012:H3a-H4a} and GST-4~\cite{Model:2013:GST}, are considered in addition to the nominal flux of H3a~\cite{Model:2012:H3a-H4a}. The PPMs are obtained using the alternative fluxes, and the total variation is shown by the orange band in Fig.~\ref{fig:passing_proba_systematics}. Although not shown, the models obtained from H3a and H4a are similar, as expected because these two fluxes differ only in the extra-galactic components, where the H4a model assumes all protons. So, the PPM obtained from H4a has slightly larger values at high stochasticity. The model obtained assuming the GST-4 flux has larger deviations from the nominal PPM, as it introduces a fourth extra-galactic component.

Since snow accumulation occurs throughout the year (albeit not evenly), parameterizing it by discrete years introduces small mismatches.   First, inefficiency contours with a value of 0.01 are obtained using different years of data. The baseline inefficiency contours (modeled with the full data) with values of 0.0093 and 0.0106 are then obtained. The values are chosen such that the spacing between these two contours is comparable to the variation of the yearly contours between adjacent years. Subsequently, the PPMs are remodeled after applying baseline inefficiency cuts at 0.0093 and 0.0106 to both data and MC, respectively. The resultant uncertainty are shown by the blue band in Fig.~\ref{fig:passing_proba_systematics}. Applying a looser inefficiency cut, \ie, 0.0106, leads to more events passing through the veto, raising the modeled passing probability curve.

These systematic effects on the modeling of the PPM are summarized in Fig.~\ref{fig:passing_proba_systematics}. The nominal PPM, $\mathcal{P}_{\mathrm{nominal}}(s)$, is shown by the solid black curve. The singleness threshold has the largest systematic effect, followed by the short-term snow effect. The total systematic uncertainty per stochasticity bin ($\sigma^{+}(s)$ or $\sigma^{-}(s)$, depending on the direction) is obtained by summing individual systematic uncertainties in quadrature, assuming that these effects are independent. The two dashed black curves represent PPMs incorporating the total systematic uncertainty for each stochasticity bin: $\mathcal{P}_{\mathrm{syst}}(s) = \mathcal{P}_{\mathrm{nominal}}(s)^{+\sigma^{+}(s)}_{-\sigma^{-}(s)}$. We then calculate the total systematic uncertainty associated with the nominal background events in the signal region using $\mathcal{P}_{\mathrm{syst}}(s)$, following the same procedure in Section~\ref{subsec:estimated-background}. The final result is $1.97^{+0.12}_{-0.14}$.

The background normalization for the downgoing sample, denoted by $\muonNormDPeV$, is assigned a Gaussian prior with a nominal value of 1.97.  We conservatively assume a prior width of $\sqrt{1.97}$. 
%
The upper bound on $\muonNormDPeV$ is set to $1.97 + 2 \times \sqrt{1.97} = 4.79$, two standard deviations from the nominal value. The lower bound is set to 0 to ensure a physical muon background flux.

\section{Correlation coefficients}
\label{app:correlations}

\begin{figure*}[h]
\includegraphics[width=0.49\textwidth]{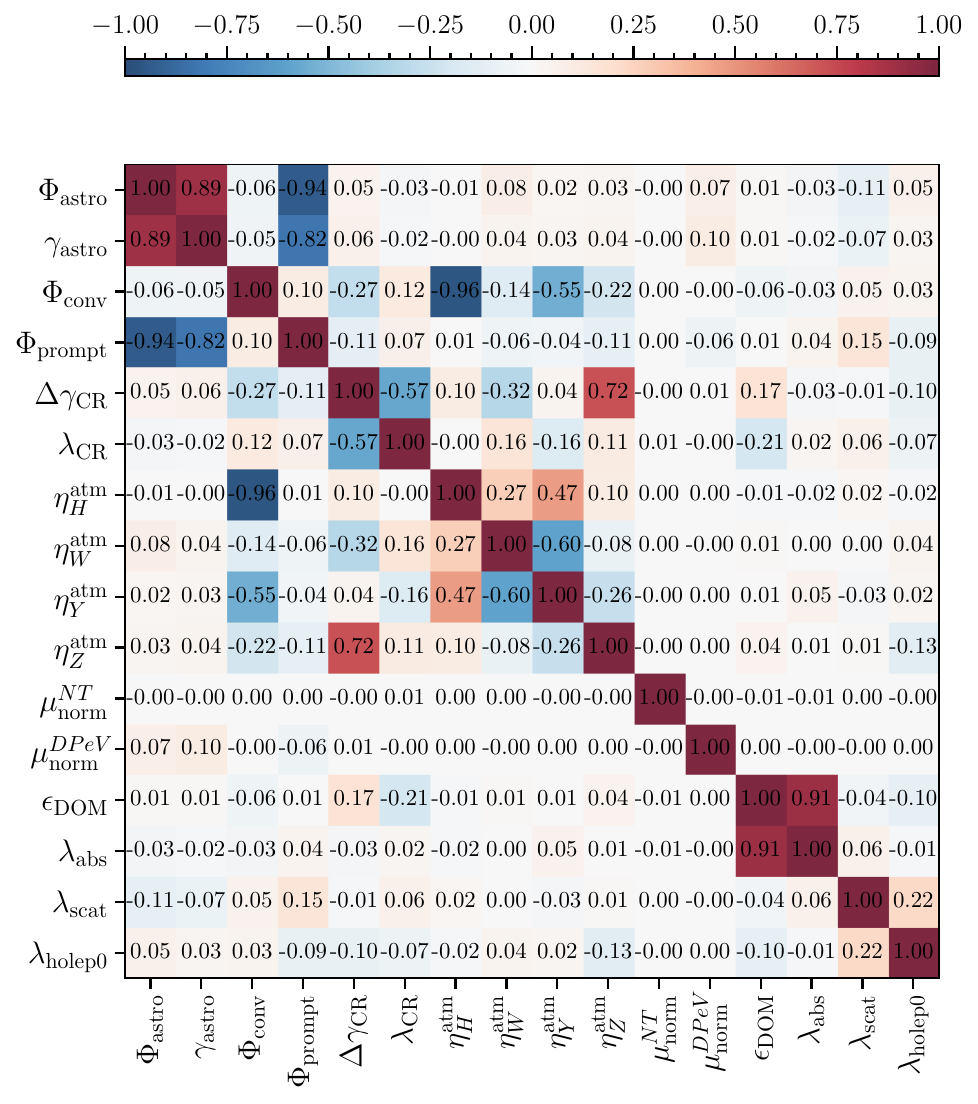}
\includegraphics[width=0.49\textwidth]{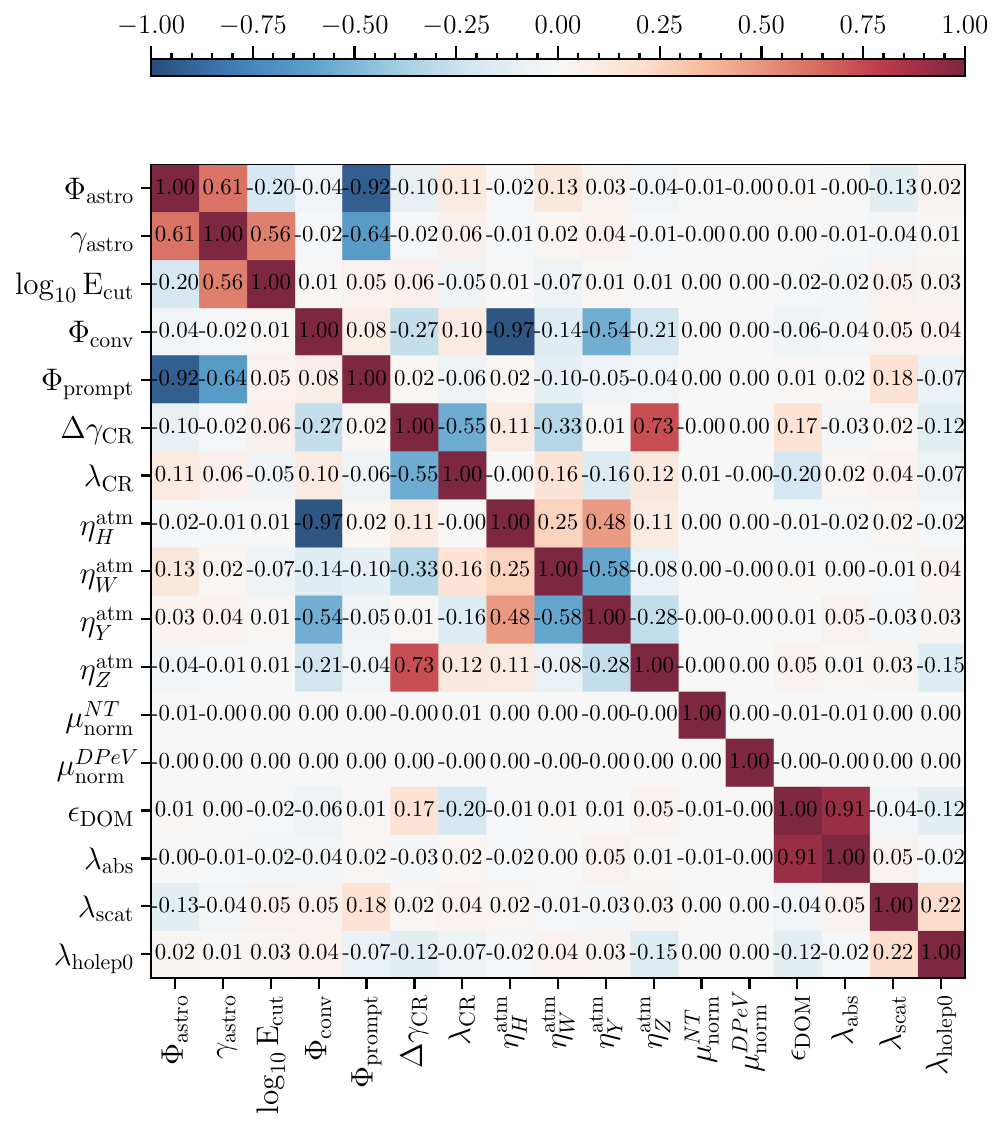}
    \caption{Correlation coefficients between all parameters from the SPL and SPL+cutoff fits.}
    \label{fig:corr_matrix}
\end{figure*}

After obtaining the best-fit parameters, the correlation matrix of all parameters could be calculated using the Fisher information matrix, which can be approximated by the Hessian matrix (second derivative of the log-likelihood) computed at the best-fit point. Fig.~\ref{fig:corr_matrix} shows the correlation matrices for the SPL and SPL+cutoff parameters.

\section{Astrophysical source searches}

\begin{figure*}[h]
\includegraphics[width=0.45\textwidth]{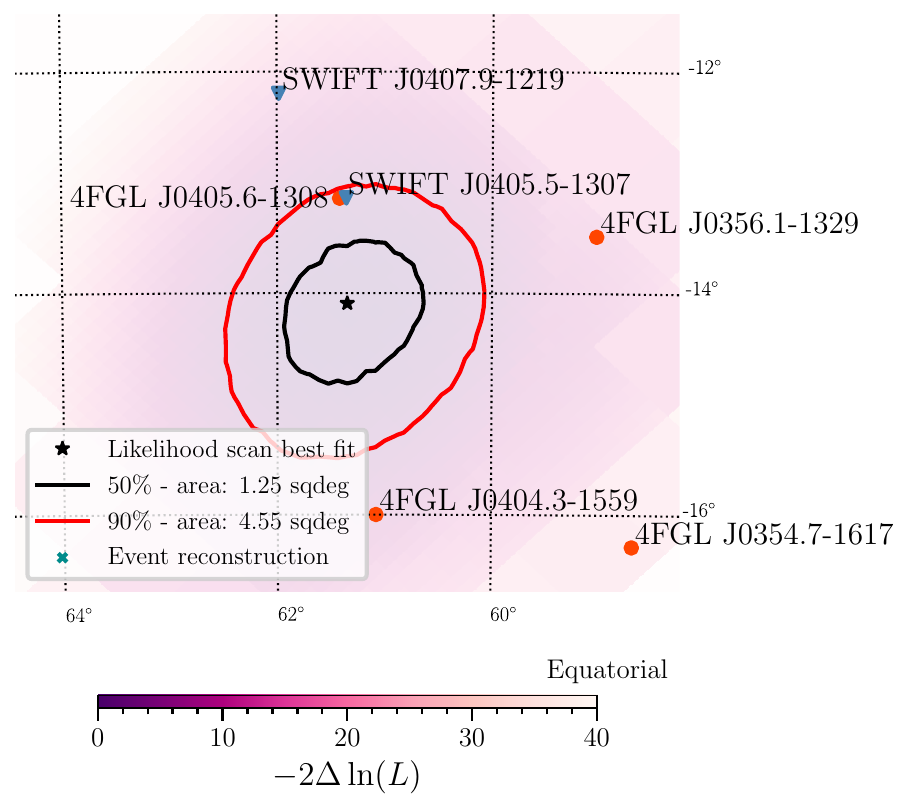}
\includegraphics[width=0.5\textwidth]{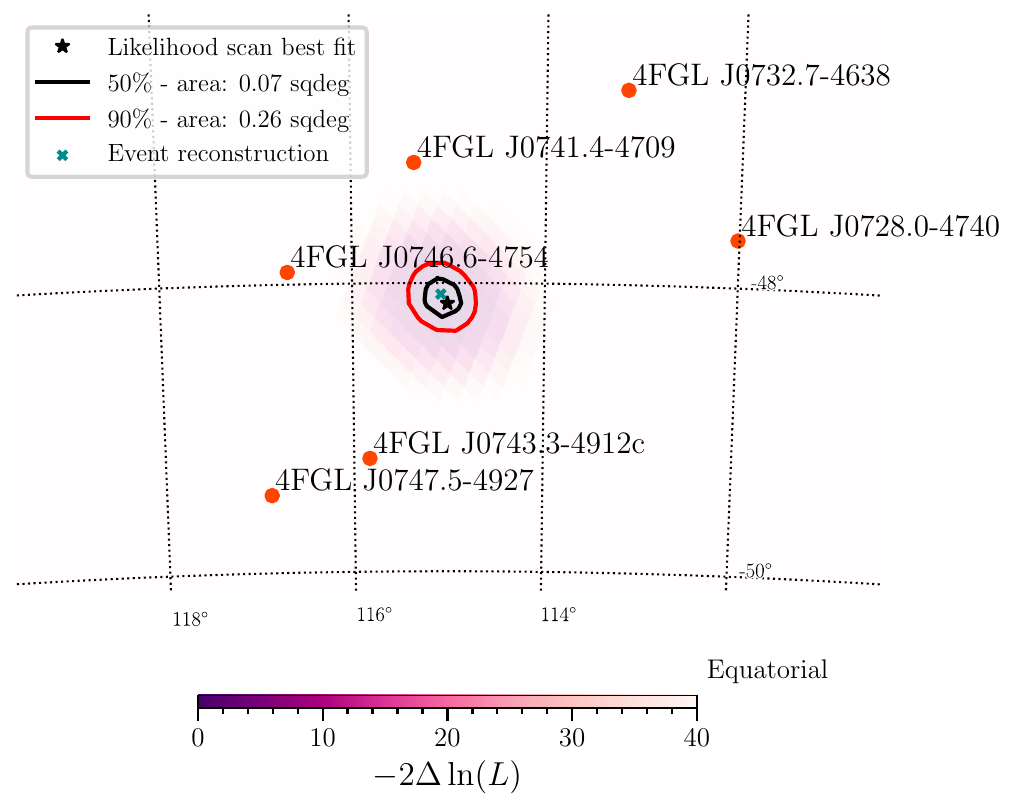}
    \caption{Angular uncertainty contours and nearby high-energy astrophysical sources for the two observed events. The contours are based on a likelihood scan~\cite{PS:2023:IceCat-1}. The first event (left) has one associated source, which is reported in both the 4FGL and SWIFT catalogs. The counterpart is PKS 0403-13, a BL Lac object at a redshift of 0.57. The second event (right) does not have associated sources.}
    \label{fig:event_scan}
\end{figure*}

To perform the point source search using high-energy catalogs, the locations of the astrophysical sources and the directional uncertainty contours for each event are compared. The results are shown in Fig.~\ref{fig:event_scan}.

\clearpage

\end{document}